\newif\ifdraft
\draftfalse

\newif\ifcameraready
\camerareadytrue

\documentclass[camera,letterpaper,nomarginnotes,nonarrowgutter]{jpaper}
\newcounter{version}
\usepackage{mathptmx} 
\usepackage[sort,compress]{cite}
\usepackage{amsmath,amssymb,amsfonts}
\usepackage{algorithmic}
\usepackage{graphicx}
\usepackage{textcomp}
\usepackage{xcolor}
\usepackage{balance}
\usepackage{amsmath,amssymb,amsfonts}
\usepackage{graphicx}
\usepackage{textcomp}
\usepackage{pifont}
\usepackage{booktabs}
\usepackage{glossaries} 
\usepackage{setspace}
\usepackage{listings}
\usepackage[linesnumbered]{algorithm2e}
\usepackage{siunitx} 
\usepackage{subcaption}
\usepackage{tikz}
\usepackage{xcolor}
\usepackage{multirow}
\usepackage{enumitem}

\usepackage{titlesec}

\usepackage{interval} 
\usepackage{duckuments} 
\usepackage{algorithmic} 

\usepackage[us,12hr]{datetime}
\usepackage[en-GB, useregional=numeric]{datetime2}
\usepackage{fourier-orns}
\usepackage{fancyhdr}
\usepackage{pgfornament}
\usetikzlibrary{calc}

\usepackage{url}

\definecolor{gfored}{rgb}{0.580, 0.050, 0.211}
\definecolor{ao}{rgb}{0.007, 0.520, 0.867}
\definecolor{moegi}{rgb}{0.357, 0.537, 0.188}
\definecolor{jl}{rgb}{1.0, 0.2, 0.8}
\definecolor{brown(web)}{rgb}{0.65, 0.16, 0.16}
\definecolor{bisque}{rgb}{1.0, 0.89, 0.77}
\definecolor{nbs}{rgb}{0.88, 0.07, 0.37}
\definecolor{yt}{rgb}{0.58, 0.44, 0.86}
\definecolor{iy}{rgb}{0.0, 0.36, 0.05}
\definecolor{mel}{rgb}{0.9, 0.55, 0.31}

\newcommand{\dingOne}{\circledtest{1}}
\newcommand{\dingTwo}{\circledtest{2}}
\newcommand{\dingThree}{\circledtest{3}}
\newcommand{\dingFour}{\circledtest{4}}
\newcommand{\dingFive}{\circledtest{5}}
\newcommand{\dingSix}{\circledtest{6}}

\newcommand{\one}{1)}
\newcommand{\two}{2)}
\newcommand{\three}{3)}
\newcommand{\four}{4)}
\newcommand{\five}{5)}
\newcommand{\six}{6)}

\newcommand{\head}[1]{\noindent\textbf{#1.}}

\newcommand{\xxx}[1]{\param{XXX}} 

\newcommand{\ignore}[1]{}

\newcommand{\param}[1]{\textcolor{red}{#1}}

\usepackage[colorinlistoftodos,prependcaption,textsize=scriptsize]{todonotes} 

\usepackage{marginnote} 
\let\marginpar\marginnote
\ifdraft
    \usepackage[colorinlistoftodos,prependcaption,textsize=tiny]{todonotes}
    \usepackage{color,soul}
    \paperwidth=\dimexpr \paperwidth + 4cm\relax
    \oddsidemargin=\dimexpr\oddsidemargin + 2cm\relax
    \evensidemargin=\dimexpr\evensidemargin + 2cm\relax
    \marginparwidth=\dimexpr \marginparwidth + 2cm\relax

    \newcommand{\agy}[1]{\textcolor{gfored}{#1}}
    \newcommand{\agycomment}[1]{\todo[size=\scriptsize, linecolor=orange, bordercolor=orange, backgroundcolor=white]{\textcolor{gfored}{\textbf{@gy:} #1}}}
    \renewcommand{\agycomment}[1]{{\textcolor{gfored}{\textbf{\hl{[@gy:}} \hl{#1}\textbf{]}}}}

    \newcommand{\mscomment}[1]{\todo[size=\scriptsize, linecolor=orange, bordercolor=orange, backgroundcolor=white]{\textcolor{red}{\textbf{@MS:} #1}}}
    \renewcommand{\mscomment}[1]{{\textcolor{red}{\textbf{\hl{[@MS:}} \hl{#1}\textbf{]}}}}

    \newcommand{\atb}[1]{\textcolor{ao}{#1}}

    \newcommand{\atbcomment}[1]{\todo[size=\scriptsize, linecolor=orange, bordercolor=orange, backgroundcolor=white]{\textcolor{ao}{\textbf{@atb:} #1}}}

    \newcommand{\yct}[1]{\textcolor{yt}{#1}}
    \newcommand{\yctcomment}[1]{\todo[size=\scriptsize, linecolor=orange, bordercolor=orange, backgroundcolor=white]{\textcolor{yt}{\textbf{@yct:} #1}}}

    \newcommand{\gf}[1]{\textcolor{blue}{#1}}
    \newcommand{\gfcomment}[1]{\todo[size=\scriptsize, linecolor=orange, bordercolor=orange, backgroundcolor=white]{\textcolor{blue}{\textbf{@gf:} #1}}}
    
    \newcommand{\nb}[1]{\textcolor{nbs}{#1}}
    \newcommand{\nbcomment}[1]{\todo[size=\scriptsize, linecolor=orange, bordercolor=orange, backgroundcolor=white]{\textcolor{nbs}{\textbf{@nb:} #1}}}

    \newcommand{\hluo}[1]{\textcolor{moegi}{#1}}
    \newcommand{\hluocomment}[1]{\todo[size=\scriptsize, linecolor=orange, bordercolor=orange, backgroundcolor=white]{\textcolor{moegi}{\textbf{@hluo:} #1}}}

    \newcommand{\mel}[1]{\textcolor{mel}{#1}}
    \newcommand{\melcomment}[1]{\todo[size=\scriptsize, linecolor=red, bordercolor=red, backgroundcolor=white]{\textcolor{mel}{\textbf{@mel:} #1}}}

    \newcommand{\iey}[1]{\textcolor{iy}{#1}}
    \newcommand{\ieycomment}[1]{\todo[size=\scriptsize, linecolor=orange, bordercolor=orange, backgroundcolor=white]{\textcolor{iy}{\textbf{@iey:} #1}}}

    \newcommand{\omcomment}[1]{\todo[size=\scriptsize, linecolor=orange, bordercolor=orange, backgroundcolor=white]{\textcolor{teal}{\textbf{@ste:} #1}}}

\else
    \renewcommand{\param}[1]{\textcolor{black}{#1}}

    \newcommand{\agy}[1]{{#1}}
    \newcommand{\agycomment}[1]{}
    \newcommand{\agyinline}[1]{}

    \newcommand{\mscomment}[1]{}
    
    \newcommand{\atb}[1]{{#1}}
    \newcommand{\atbcomment}[1]{}
    
    \newcommand{\yct}[1]{{#1}}
    \newcommand{\yctcomment}[1]{}
    
    \newcommand{\gf}[1]{{#1}}
    \newcommand{\gfcomment}[1]{}

    \newcommand{\nb}[1]{{#1}}
    \newcommand{\nbcomment}[1]{}

    \newcommand{\hluo}[1]{{#1}}
    \newcommand{\hluocomment}[1]{}

    \newcommand{\mel}[1]{{#1}}
    \newcommand{\melcomment}[1]{}

    \newcommand{\iey}[1]{#1}
    \newcommand{\ieycomment}[1]{}

    \newcommand{\omcomment}[1]{}

\fi

\newcommand*\nCHIPS{120}
\newcommand*\nMODULES{18}

\definecolor{frenchblue}{rgb}{0.19, 0.55, 0.91}

\usepackage[most]{tcolorbox} 
\tcbset{before skip=4.1pt, after skip=4.1pt}

\newtcolorbox[auto counter]{obsx}[3][]{%
    colframe = #2!45,
    colback  = #2!10,
    coltitle = #2!20!black, 
    colbacktitle=#2!20,
    coltitle=black,
    fonttitle=\bfseries, 
    title=#3~\thetcbcounter.\ ,
    enhanced,
    attach boxed title to top left={yshift=-2.8mm, xshift=0.15cm},
    bottom=-2.2pt,
    #1%
}

\usepackage[most]{tcolorbox} 
\newtcolorbox[auto counter]{tkx}[2][]{%
    enhanced, breakable, center title,
    colframe = #2!45,
    colback  = #2!10,
    colbacktitle=#2!20,
    left=-0.5pt,
    right=-0.5pt,
    bottom=-2pt,
    top=-0.25pt,
    #1%
}
\newcounter{obs}
\newcommand\observation[1]{
\refstepcounter{obs}
\begin{tkx}{gray}
\noindent\textbf{Observation~\theobs.} #1
\end{tkx}
}

\newcounter{tkw}
\newcommand\takeaway[1]{
\stepcounter{tkw}
\begin{tkx}{iy}
\noindent\textbf{Takeaway~\thetkw.} #1
\end{tkx}
}

\newcommand{\dataMovementProblemsCitations}[0]{\cite{
mutlu2013memory,
mutlu2015research,
dean2013tail,
kanev_isca2015,
ferdman2012clearing,
wang2014bigdatabench,
mutlu2019enabling,
mutlu2019processing,
mutlu2020intelligent,
ghose.ibmjrd19,
mutlu2020modern,
oliveira2021damov,
boroumand2018google, 
boroumand2021google,
wang2016reducing, 
pandiyan2014quantifying,
koppula2019eden,
kang2014co,
mckee2004reflections,
wilkes2001memory,
kim2012case,
wulf1995hitting,
ghose.sigmetrics20,
ahn2015scalable,
PEI,
hsieh2016transparent,
wang2020figaro,
sites1996
}}

\newcommand\pimshort{\cite{devaux2019true,ghiasi2022genstore,gomez2021benchmarkingcut,gomezluna2021benchmarking,gomez2022benchmarking,syncron,singh2020nero,skhynixpim,ke2021near,giannoula2022sparsep,denzler2021casper,IRAM_Micro_1997,C_RAM_1999,gokhale1995processing,hall1999mapping,ahn2015scalable,boroumand2018google,boroumand2016lazypim, boroumand2017lazypim, zhang2014top, kim2018grim,RVU, NIM, PEI,kim2016neurocube, boroumand2019conda, hsieh2016transparent, hsieh2016accelerating,boroumand2021mitigating,boroumand2021google,boroumand2022polynesia,boroumand2021polynesia, besta2021sisa,fernandez2020natsa,singh2019napel,lee2021hardware,kim2017grim,santos2018processing,Chi2016, Shafiee2016, seshadri2017ambit, seshadri2019dram, li2017drisa, seshadri2013rowclone, seshadri2016processing, deng2018dracc, xin2020elp2im,  song2017pipelayer,gao2019computedram, eckert2018neural, aga2017compute,dualitycache,seshadri2016buddy,seshadri.bookchapter17,seshadri2018rowclone,seshadri2015fast,li2016pinatubo,ferreira2021pluto,ferreira2022pluto,flashcosmos,truong2022adapting,truong2021racer,olgun2021quactrng,kim2018dram,bostanci2022dr,olgun2021pidram,ali2019memory,li2018scope,orosa2021codic,sharad2013ultra,
gao2021parabit,choi2020flash,han2019novel,merrikh2017high,wang2018three,cali2020genasm,nag2019gencache,kang2014energy,wang2023infinity,kang2015energy,chang2016low, hajinazarsimdram,sutradhar2021look,sutradhar2020ppim,lenjani2020fulcrum,peng2023chopper,oliveira2022accelerating,singh2021fpga,oliveira2023dappa,oliveira2022methodologies,oliveira2022heterogeneous,chen2023simplepim,gupta2023evaluating,gomez2023evaluating,oliveira2023transpimlib,diab2023framework,mao2022genpip,singh2022accelerating}}

\newcommand\pnmshort{\cite{devaux2019true,ghiasi2022genstore,gomez2021benchmarkingcut,gomezluna2021benchmarking,gomez2022benchmarking,syncron,singh2020nero,skhynixpim,ke2021near,giannoula2022sparsep,denzler2021casper,IRAM_Micro_1997,C_RAM_1999,gokhale1995processing,hall1999mapping,ahn2015scalable,boroumand2018google,boroumand2016lazypim, zhang2014top, kim2018grim, RVU, NIM, PEI,kim2016neurocube, boroumand2019conda, hsieh2016transparent, cali2020genasm,hsieh2016accelerating,boroumand2021mitigating,boroumand2021google,boroumand2022polynesia,boroumand2021polynesia,besta2021sisa,fernandez2020natsa,singh2019napel,lee2021hardware,kim2017grim,boroumand2017lazypim,santos2018processing,lenjani2020fulcrum}}

\newcommand\pumshort{\cite{Chi2016, Shafiee2016, seshadri2017ambit, seshadri2019dram, li2017drisa, seshadri2013rowclone, seshadri2016processing, deng2018dracc, xin2020elp2im, song2017pipelayer,gao2019computedram, eckert2018neural, aga2017compute,dualitycache,besta2021sisa,seshadri2016buddy,seshadri.bookchapter17,seshadri2018rowclone,seshadri2015fast,li2016pinatubo,ferreira2021pluto,ferreira2022pluto,flashcosmos,truong2022adapting,truong2021racer,olgun2021quactrng,kim2018dram,bostanci2022dr,olgun2021pidram,ali2019memory,li2018scope,orosa2021codic,sharad2013ultra,gao2021parabit,choi2020flash,han2019novel,merrikh2017high,wang2018three,nag2019gencache,kang2014energy,wang2023infinity,kang2015energy, chang2016low,hajinazarsimdram,sutradhar2021look,sutradhar2020ppim,peng2023chopper,deoliveira2024mimdram, yuksel2024functionallycomplete}}

\newcommand{\figref}[1]{Fig.~\ref{#1}}

\newcommand{\secref}[1]{§\ref{#1}}

\newcommand{\maj}[1]{\texttt{MAJ{#1}}}

\newcommand*\circledtest[1]{\tikz[baseline=(char.base)]{
            \node[shape=circle,fill,inner sep=0.3pt] (char) {\textcolor{white}{#1}};}}

\newcommand*\circledt[2]{\tikz[baseline=(char.base)]{
    \node[shape=circle, draw, fill=#1, inner sep=0.05pt] (char) {\vphantom{WAH1g}\textcolor{white}{#2}};}}

\newcommand{\dingA}{\circledt{blue}{a}}
\newcommand{\dingB}{\circledt{blue}{b}}
\newcommand{\dingC}{\circledt{blue}{c}}
\newcommand{\dingD}{\circledt{blue}{d}}
\newcommand{\dingE}{\circledt{blue}{e}}

\ifcameraready

    \newcommand{\atbcr}[2]{\ifnum#1>-1\textcolor{black}{#2}\else{#2}\fi}
    \newcommand{\ieycr}[2]{\ifnum#1>-1\textcolor{black}{#2}\else{#2}\fi}
    \newcommand{\omcr}[2]{\ifnum#1>-1\textcolor{black}{#2}\else{#2}\fi}
    \newcommand{\gfcr}[2]{\ifnum#1>-1\textcolor{black}{#2}\else{#2}\fi}

\else
    \paperwidth=\dimexpr \paperwidth + 4cm\relax
    \oddsidemargin=\dimexpr\oddsidemargin + 2cm\relax
    \evensidemargin=\dimexpr\evensidemargin + 2cm\relax
    \marginparwidth=\dimexpr \marginparwidth + 2cm\relax
    
    \newcommand{\atbcr}[2]{\ifnum#1=\value{version}\textcolor{ao}{#2}\else{#2}\fi}

    \newcommand{\ieycr}[2]{\ifnum#1=\value{version}\textcolor{blue}{#2}\else{#2}\fi}

    \newcommand{\omcr}[2]{\ifnum#1=\value{version}\textcolor{blue}{#2}\else{#2}\fi}
    \newcommand{\gfcr}[2]{\ifnum#1=\value{version}\textcolor{red}{#2}\else{#2}\fi}

\fi

\makeatletter
\g@addto@macro{\normalsize}{%
 \setlength{\abovedisplayskip}{2pt plus 1pt minus 1pt}
 \setlength{\belowdisplayskip}{2pt plus 1pt minus 1pt}
  \setlength{\abovedisplayshortskip}{0pt}
  \setlength{\belowdisplayshortskip}{0pt}
  \setlength{\intextsep}{2pt plus 1pt minus 1pt}
  \setlength{\textfloatsep}{2pt plus 1pt minus 1pt}
  \setlength{\skip\footins}{5pt plus 1pt minus 1pt}}
\makeatother

\usepackage{tikz}
\usetikzlibrary{arrows}

\newcommand{\tras}[0]{t_{RAS}}
\newcommand{\trp}[0]{t_{RP}}
\newcommand{\trc}[0]{t_{RC}}
\newcommand{\trefi}[0]{t_{REFI}}
\newcommand{\trefw}[0]{t_{REFW}}
\newcommand{\trfc}[0]{t_{RFC}}
\newcommand{\trrd}[0]{t_{RRD}}

\newcommand{\act}[0]{\texttt{ACT}}
\newcommand{\pre}[0]{\texttt{PRE}}
\newcommand{\refresh}[0]{REF}
\newcommand{\wri}[0]{\texttt{WR}}
\newcommand{\rd}[0]{\texttt{RD}}

\newcommand{\apaLong}[0]{\act{} $\rightarrow$ \pre{} $\rightarrow$ \act{}}

\newcommand{\apa}[0]{\texttt{APA}}

\newcommand{\rfirst}[0]{$R_{F}$}
\newcommand{\rsecond}[0]{$R_{S}$}

\newcommand{\vddh}[0]{\texttt{$V_{DD}/2$}}
\newcommand{\vdd}[0]{\texttt{$V_{DD}$}}
\newcommand{\vpp}[0]{\texttt{$V_{PP}$}}

\newcommand{\apaTime}[0]{{\act{} \rfirst{} $\xrightarrow[]{t_{1}}$ \pre{} $\xrightarrow[]{t_{2}}$ \act{} \rsecond{}}}
\newcommand{\apaex}[0]{{\act{} 0 $\rightarrow$ \pre{} $\rightarrow$ \act{} 7}}

\newcommand{\pum}[0]{PUM}
\newcommand{\pim}[0]{PIM}
\newcommand{\pnm}[0]{PNM}

\newcommand{\pud}[0]{{PUD}}

\usepackage[shortcuts]{extdash}
\newcommand{\mrc}[0]{\texttt{Multi\-/RowCopy}}
\newacronym{iqr}{$IQR$}{inter-quartile range}
\newacronym{act}{\act{}}{activate}
\newacronym{pre}{\pre{}}{precharge}
\newacronym{ref}{\refresh{}}{refresh}
\newacronym{wr}{\wri{}}{write}
\newacronym{rd}{\rd{}}{read}
\newacronym{pim}{\pim{}}{Processing-In-Memory}
\newacronym{pnm}{\pnm{}}{Processing-Near-Memory}

\newacronym{pum}{\pum{}}{Processing-Using-Memory}
\newacronym{pud}{\pud{}}{Processing-Using-DRAM}
\newacronym{apa}{\apa{}}{\act{} $\rightarrow$ \pre{} $\rightarrow$ \act{}}
\newacronym{jedec}{JEDEC}{Joint Electron Device Engineering Council}

\newacronym{trefw}{$\trefw$}{refresh window}

\newacronym{tras}{$\tras$}{the latency of sensing the row's data and fully restoring a DRAM cell's charge}
\newacronym{trp}{$\trp$}{the latency of de-asserting a wordline and precharging the bitlines to \vddh{}}
\newacronym{trc}{$\trc$}{the minimum time needed between two consecutive row activations targeting the same bank}
\newacronym{trefi}{$\trefi$}{refresh interval}
\newacronym{trfc}{$\trfc$}{refresh latency}
\newacronym{trrd}{$\trrd$}{the minimum time needed between two consecutive row activations targeting the same rank}
\newacronym{puf}{PUF}{physical unclonable function}
\newacronym{trn}{TRN}{true random number}

\usepackage[bookmarks=true,breaklinks=true,colorlinks,linkcolor=black,citecolor=NavyBlue,urlcolor=blue]{hyperref}

\def\BibTeX{{\rm B\kern-.05em{\sc i\kern-.025em b}\kern-.08em
    T\kern-.1667em\lower.7ex\hbox{E}\kern-.125emX}}
    
\newcommand{\versionnum}[0]{1.0}

\ifcameraready
    \fancypagestyle{plain}{
    \fancyhead{}
    }
\else
    \fancypagestyle{firstpage}
    {
    }
\fi

\makeatletter
\def\bstctlcite{\@ifnextchar[{\@bstctlcite}{\@bstctlcite[@auxout]}}
\def\@bstctlcite[#1]#2{\@bsphack
  \@for\@citeb:=#2\do{%
    \edef\@citeb{\expandafter\@firstofone\@citeb}%
    \if@filesw\immediate\write\csname #1\endcsname{\string\citation{\@citeb}}\fi}%
  \@esphack}
\makeatother 

\usepackage[available,functional]{usenixbadges}

\newcommand{\affilETH}{$^1$}
\newcommand{\affilYahya}{$^{1,2}$}
\newcommand{\affilOnur}{$^{1}$}

\begin{document}
\bstctlcite{IEEEexample:BSTcontrol}

\title{Simultaneous Many-Row Activation in \ieycr{0}{Off-the-Shelf} DRAM \ieycr{0}{Chips}:\\Experimental Characterization and Analysis}

\author{
{\.I}smail~Emir~Y{\"u}ksel\affilETH{}\quad
Yahya~Can~Tu\u{g}rul\affilYahya{}\quad
F.~Nisa~Bostanc{\i}\affilETH{}\quad
Geraldo~F.~Oliveira\affilETH{}\\
A.~Giray~Ya\u{g}l{\i}k\c{c}{\i}\affilETH{}\quad
Ataberk~Olgun\affilETH{}\quad
Melina~Soysal\affilETH{}\quad
Haocong~Luo\affilETH{}\\
Juan~Gómez-Luna\affilETH{}\quad
Mohammad~Sadrosadati\affilETH{}~\quad
Onur~Mutlu\affilOnur{}\vspace{1.5pt}\\
$^1$\emph{ETH~Z{\"u}rich}~\qquad
$^2$\emph{TOBB University of Economics and Technology}}

\maketitle

\pagestyle{plain}

\setcounter{version}{3}
\begin{abstract}
We experimentally analyze the computational capability of \omcr{1}{commercial off-the-shelf (COTS)} DRAM chips and the robustness of these capabilities under various timing delays between DRAM commands, data patterns, \gfcr{1}{temperature}, and voltage levels. 
We extensively characterize \nCHIPS{} \omcr{1}{COTS} DDR4 chips from two major manufacturers. 
We highlight \omcr{2}{four} key results of our study.~First, \omcr{1}{COTS} DRAM chips are capable of \one{} simultaneously activating up to 32 rows~\ieycr{1}{(i.e., simultaneous many-row activation)}, \two{} executing \ieycr{1}{a majority of X (\maj{X}) operation where X>3 (i.e., } \maj{5}, \maj{7}, and \maj{9}~\ieycr{1}{operations)}, and \three{} copying a DRAM row (concurrently) to up to 31 other DRAM rows, \ieycr{1}{which we call \mrc{}}. 
Second, storing multiple copies of \ieycr{1}{\maj{X}'s input operands} on all simultaneously activated rows drastically increases the success rate (i.e., the percentage of DRAM cells that \emph{correctly} perform the computation) \omcr{2}{of the \maj{X} operation}.
For example, \maj{3} with 32-row activation \ieycr{0}{(i.e., replicating each \maj{3}'s input \ieycr{1}{operands} 10 times)} has \ieycr{2}{a} \param{30.81}\% higher average success rate than \maj{3} with 4-row activation \ieycr{0}{(i.e., no replication)}. 
Third, data pattern affect\omcr{1}{s} the success rate of \maj{X} and \ieycr{1}{\mrc{}} operations by \param{11.52}\% and \param{0.07}\% on average.~\ieycr{1}{Fourth, simultaneous many-row activation, \maj{X}\gfcr{1}{,} and \mrc{} operations are highly resilient to temperature and voltage changes, with small success rate variations of at most 2.13\% among all tested operations. We believe these empirical results demonstrate the promising potential of using DRAM as a computation substrate. To aid future
research and development, we open-source our infrastructure at \url{https://github.com/CMU-SAFARI/SiMRA-DRAM}.}
\end{abstract}

\glsresetall
\glsresetall
\setcounter{version}{3}

\section{Introduction}

Modern computing systems \atb{move vast amounts of data} between main memory (DRAM) and \ieycr{0}{processing elements} (e.g., CPU\ieycr{0}{, GPU, TPU, and FPGA})~\cite{mutlu2019processing, mutlu2020modern}. Unfortunately, this data movement is a major bottleneck that consumes a large fraction of execution time and energy \gf{in many modern applications}\dataMovementProblemsCitations{}. 
To address this problem, \gls{pim}~\pimshort{} \omcr{1}{is} a promising paradigm to alleviate data movement bottlenecks~\cite{seshadri2017ambit,seshadri2015fast,seshadri2019dram,seshadri2016processing,seshadri2017simple,seshadri2016buddy,li2016pinatubo,xin2020elp2im,angizi2019redram}.~There are two main approaches to \gls{pim}~\cite{ghose.ibmjrd19, mutlu2020modern}: \one{}~\gls{pnm}~\pnmshort{}, where computation logic is added near the memory arrays (e.g., in a DRAM chip or at the logic layer of a 3D-stacked memory~\cite{HMC2, HBM,lee2016simultaneous}); and
\two{}~\gls{pum}~\pumshort{}, where computation is performed by \ieycr{0}{leveraging} the analog operational properties of the memory circuitry.

A subset of \gls{pim} proposals devise mechanisms that enable \gls{pum} using DRAM cells for computation, including data copy and initialization~\cite{seshadri.bookchapter17, seshadri2013rowclone,seshadri2018rowclone,chang2016low, gao2019computedram,olgun2021pidram, deoliveira2024mimdram}, Boolean logic~\cite{seshadri2017ambit,seshadri2015fast,seshadri2019dram,seshadri2016processing,seshadri2017simple,seshadri2016buddy,gao2019computedram,gao2022frac,xin2020elp2im,besta2021sisa, li2017drisa, deoliveira2024mimdram, yuksel2024functionallycomplete}, majority-based arithmetic~\cite{hajinazar2020simdram, gao2019computedram, seshadri2017ambit, deng2018dracc, li2017drisa, angizi2019graphide, li2018scope, deoliveira2024mimdram}, and \ieycr{1}{lookup table based operations~\cite{ferreira2022pluto,deng2019lacc,sutradhar2021look,sutradhar2020ppim}}. We refer to DRAM-based \gls{pum} as \omcr{2}{\emph{\gls{pud}}} and the computation performed using DRAM cells as \gls{pud} operations.

\gls{pud} benefits from the bulk data parallelism in DRAM devices to perform bulk bitwise \gls{pud} operations. Prior works show that bulk bitwise operations are used in a wide variety of important applications, including databases and web search~\cite{chan1998bitmap,o2007bitmap,li2014widetable,li2013bitweaving,goodwin2017bitfunnel,seshadri2013rowclone,seshadri2017ambit,seshadri2015fast,hajinazar2020simdram,wu2005fastbit,wu1998encoded,redis-bitmaps}, data analytics~\cite{perach2023understanding,seshadri2017ambit,jun2015bluedbm,torabzadehkashi2019catalina,lee2020smartssd}, graph processing~\cite{beamer2012direction,besta2021sisa,li2016pinatubo,gao2021parabit,hajinazar2020simdram}, genome analysis~\cite{alser2017gatekeeper,loving2014bitpal,xin2015shifted,cali2020genasm,kim2017grim,myers1999fast}, cryptography~\cite{han1999optical,tuyls2005xor}, set operations~\cite{besta2021sisa,seshadri2017ambit}, and hyper-dimensional computing~\cite{kanerva1992sparse,kanerva2009hyperdimensional,karunaratne2020memory}.

Recent works\gf{~\cite{gao2019computedram,gao2022frac, olgun2023dram,olgun2021quactrng,olgun2021pidram}} experimentally demonstrate that some of these \gls{pud} operations can be realized in \omcr{1}{commercial} off-the-shelf \omcr{1}{(COTS)} DRAM chips by operating beyond manufacturer-recommended DRAM timing parameters. 
To do so, \omcr{2}{these} works~\cite{gao2019computedram,gao2022frac, olgun2023dram,olgun2021quactrng,olgun2021pidram} carefully \mel{engineer} a sequence of DRAM commands \gf{that} allows the DRAM chip to \gf{activate} (i.e., open) multiple (e.g., 2 or 4) DRAM rows \gf{\emph{simultaneously}} depending on the DRAM row addresses\gf{, \omcr{2}{via} a process that we refer to as \ieycr{3}{\emph{multiple row activation}.}} 
By performing \ieycr{3}{multiple row activation}, prior works can 
1) copy \gf{data} between two DRAM rows~\cite{gao2019computedram,olgun2021pidram}, 
2) \gf{perform} three-input majority \gf{computation}~\cite{gao2019computedram,gao2022frac, olgun2023dram}, and
3) \gf{generate} true-random numbers~\cite{olgun2021quactrng,olgun2021pidram}. 
To investigate the effectiveness of \gls{pud} operations in \ieycr{1}{COTS} DRAM chips, such works perform extensive characterization of real DDR3 and DDR4 chips to identify the appropriate timing delays between DRAM commands that lead to \gls{pud} operations.

Even though prior works show that \ieycr{1}{COTS} DRAM chips can perform \gls{pud} operations, there are several questions about the \emph{effectiveness} and \emph{robustness} of \gls{pud} operations \gf{in \ieycr{1}{COTS} DRAM chips} that should be answered \iey{to develop a rigorous understanding of the computational capabilities of modern DRAM chips}, including: 

\head{Q1} In a DRAM \ieycr{2}{subarray} with many DRAM rows (512--1024), \ieycr{3}{is it possible to \emph{robustly} perform simultaneous activation of more than four DRAM rows (i.e., simultaneous many-row activation)?}

\head{Q2} What other \gls{pud} operations can be realized in \ieycr{1}{COTS} DRAM chips by leveraging simultaneous many-row activation?

\head{Q3} \omcr{2}{How robustly can} \gls{pud} operations using simultaneous many-row activation \omcr{2}{be performed in COTS DRAM chips}?

\head{Q4} Can the \omcr{2}{robustness} of \gls{pud} operations be improved?

\head{Q5} What are the effects of various \gf{DRAM} operating conditions (i.e., voltage and temperature scaling, data pattern dependenc\omcr{3}{e}, and \gf{different} timing delays between DRAM commands) on the \omcr{2}{robustness} of \gls{pud} operations?

Our \emph{goal} is to experimentally analyze the computational capability of \ieycr{1}{COTS} DRAM chips and the robustness of \omcr{2}{such} capabilit\omcr{2}{y} \atb{under} various operating conditions by answering the \param{five} key questions above.
To this end, we conduct real DRAM chip experiments on \nCHIPS{} \ieycr{1}{COTS} DDR4 chips from two major \gf{DRAM} manufacturers contained within \nMODULES{} DRAM modules \omcr{2}{and back up our results with circuit-level simulations.} Based on \ieycr{2}{our real DRAM chip experiments, we make 18 new empirical observations and share 7 key takeaway lessons that provide answers to the five key questions above.}

\textit{\underline{Answering Q1.}} \ieycr{1}{COTS} DRAM chips \omcr{2}{are capable of} activat\omcr{2}{ing} up to 32 DRAM rows, \ieycr{3}{which we call} \gf{\emph{simultaneous many-row activation}}. 
We observe that carefully crafted DRAM commands simultaneously activate 2, 4, 8, 16, and 32 rows \omcr{2}{(\secref{sec:sim-act})}. \gf{We hypothesize that the hierarchical row decoder design\gfcr{1}{,} present in high-performance and high-density DRAM chips\gfcr{1}{,} is the primary reason behind the simultaneous many-row activation phenomenon we observe in \ieycr{1}{COTS} DRAM chips. 
Based on this hypothesis and the obtained mappings between the target row address and the observed activated DRAM rows, we derive a row decoder circuitry \omcr{2}{design} that allows simultaneous many-row activation \omcr{2}{(\secref{sec:reverse_engineering})}.}

\textit{\underline{Answering Q2.}} \gf{We observe that simultaneous many-row activation allows \ieycr{1}{COTS} DRAM chips to execute two \gls{pud} operations that prior works} \atb{do \emph{not} cover.} 
\ieycr{1}{COTS} DRAM chips are capable of  
1)~executing \atb{a majority-of-X (}\maj{X}) operation where X$>$3\atb{, i.e., \maj{5}\omcr{2}{, \maj{7}, and \maj{9} (\secref{sec:maj_char})}};
2)~copying one \atb{source} DRAM's row content \atb{concurrently} into \atb{multiple (e.g., 31) destination} DRAM row\atb{s} (\mrc{}) \omcr{2}{(\secref{sec:mrc_char})}.

\textit{\underline{Answering Q3.}} \gf{We observe that the \omcr{2}{success rate of a \gls{pud} operation (i.e.,} \gfcr{1}{the} percentage of DRAM cells that \emph{reliably} and \emph{correctly} perform \gls{pud} operation\omcr{2}{)}} \emph{significantly} varies \ieycr{2}{across \gls{pud} operations}. \gf{Our analysis shows that}  
1)~\maj{3}, \maj{5}, \maj{7}, and \maj{9} operations achieve \param{99.00}\%, \param{79.64}\%, \param{33.87}\%, and \param{5.91}\% average success rate, respectively~\omcr{2}{(\secref{sec:maj_char})}; and 
2)~copying one row's content to 1, 3, 7, 15, and 31 DRAM rows have \omcr{2}{an} average success rate of \param{99.99}\ieycr{2}{6}\%, \param{99.98}\ieycr{2}{9}\%, \param{99.99}\ieycr{2}{8}\%, \param{99.99}\ieycr{2}{9}\%, and \param{99.98}\ieycr{2}{2}\%~\omcr{2}{(\secref{sec:mrc_char})}.

\textit{\underline{Answering Q4.}} \gf{We observe that \emph{input replication} is \iey{a promising} approach to improve the \omcr{2}{success rate} of \gls{pud} operations.} 
Storing multiple copies of \maj{X}'s input \ieycr{1}{operands} on all simultaneously activated rows drastically increases the success rate. For example, our results show that performing \maj{3} with 32-row activation \omcr{2}{provides a} \param{30.81}\% \omcr{2}{higher success rate than} \omcr{3}{performing} \maj{3}  with 4-row activation~\omcr{2}{(\secref{sec:maj_char} and \secref{sec:maj_hypo})}.

\textit{\underline{Answering Q5.}} \omcr{2}{D}ata pattern \ieycr{2}{stored in simultaneously activated rows} affects the success rate of the \maj{\ieycr{2}{X}} and \mrc{} operations. We observe that data pattern affect\omcr{2}{s} the success rate by \param{11.52}\% and \param{0.07}\% on average for \maj{\ieycr{2}{X}} operations and \mrc{} operations. Temperature and voltage scaling has a \ieycr{2}{small} impact on the success rate \ieycr{2}{of simultaneous many-row activation, \maj{X}, and \mrc{} operations. Our results show that} success rate \atb{changes} by only \param{2.13}\% (\param{1.32}\%) across all tested operations when the temperature (voltage) \atb{changes from \param{\SI{50}{\celsius}} (\param{2.5V}) to \param{\SI{90}{\celsius}} (\param{2.1V})}.

\iey{By leveraging our \param{18} observations and \param{7} takeaways from \omcr{2}{extensive} experiments \omcr{2}{on real DRAM chips},} we demonstrate the performance benefits of supporting \maj{X} and \mrc{} in \ieycr{1}{COTS} DRAM chips on \one{} \param{seven} microbenchmarks and \two{} \omcr{2}{a} \iey{cold boot attack prevention mechanism}.

This paper makes the following key contributions: 
\begin{itemize}[noitemsep,topsep=0pt,parsep=0pt,partopsep=0pt,labelindent=0pt,itemindent=0pt,leftmargin=*]
\item We demonstrate, through an extensive experimental characterization of \nCHIPS{} modern DRAM chips from two major manufacturers, that modern DRAM chips can \ieycr{3}{\emph{robustly}} activate up to 32 DRAM rows \gfcr{1}{simultaneously}.
\item We demonstrate a proof-of-concept that \ieycr{3}{COTS} DRAM chips are capable of \one{} executing \maj{5}, \maj{7}, \maj{9}, and \mrc{} operations and \two{} \omcr{3}{the} increasing success rate of \maj{\ieycr{2}{X}} by replicating the input \ieycr{2}{operands} of \maj{\ieycr{2}{X}}.
\item We show the effect of DRAM operating parameters (i.e., timing delays between DRAM commands, data pattern, temperature, and voltage) on simultaneous many-row activation, \maj{\ieycr{3}{X}}, and \mrc{} operations.
\end{itemize}

We believe that our findings can be used as a basis for building new \omcr{2}{and robust} \gls{pud} mechanisms into DRAM chips and DRAM standards in the future. We hope that changes to the DRAM interface \omcr{2}{and DRAM chips, inspired by our proof-of-concept results,} can enable \gls{pud} mechanisms with lower overhead and \omcr{2}{larger performance} benefits than what we demonstrate. \ieycr{0}{To aid future research and development, we open-source our infrastructure at \url{https://github.com/CMU-SAFARI/SiMRA-DRAM}}\gfcr{1}{.}

\glsresetall
\setcounter{version}{1}
\section{Background}
\label{sec:background}

\ieycr{0}{\omcr{1}{We} briefly \omcr{1}{describe \one{}} DRAM organization, operation, timings, and \omcr{1}{\two{}} \gls{pud} in \ieycr{0}{commercial} off-the-shelf \ieycr{0}{(COTS)} DRAM chips.}
\subsection{\ieycr{0}{DRAM Organization, Operation and Timing}}

\begin{figure}[ht]
\centering
\includegraphics[width=\linewidth]{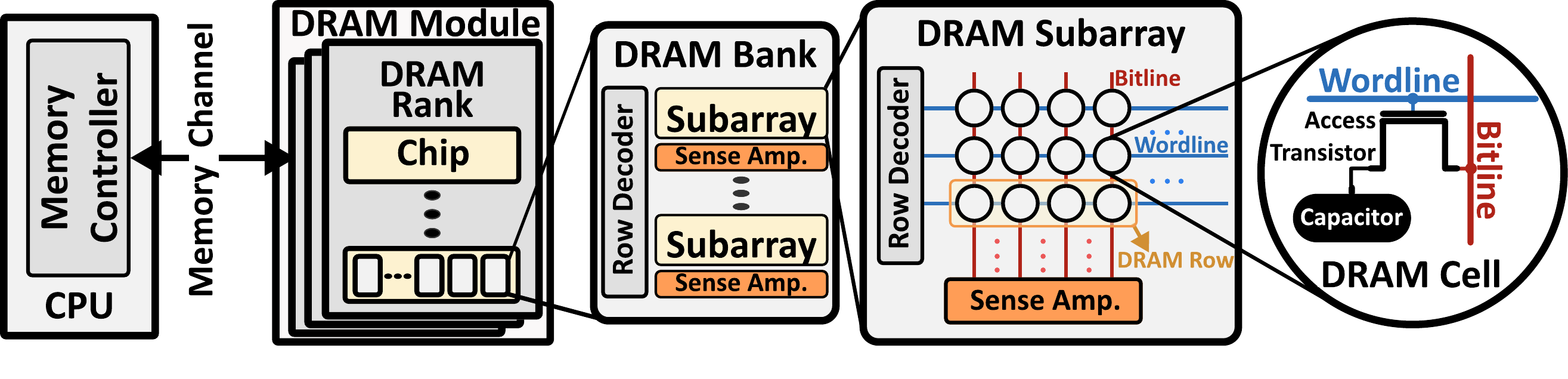}
\caption{DRAM module, chip, bank, and subarray organization.}
\label{fig:dram_organization}
\end{figure}

\head{DRAM Organization}
\figref{fig:dram_organization} shows the organization of DRAM-based memory systems. A \emph{memory channel} connects the processor (CPU) to a \emph{DRAM module}, where a module consists of multiple \emph{DRAM ranks}. A rank is formed by a set of \emph{DRAM chips} operated in lockstep. 
A DRAM chip has multiple \textit{DRAM banks}, each composed of many \emph{DRAM subarrays}.
Within a subarray, DRAM cells form a two-dimensional structure interconnected over \textit{bitlines} and \textit{wordlines}. The row decoder in a subarray decodes the row address and drives \gf{one} wordline out of many. A row of DRAM cells on the same wordline is referred to as a DRAM \omcr{1}{\emph{row}}. The DRAM cells in the same \omcr{1}{\emph{column}} are connected to the sense amplifier via a bitline.
A DRAM cell stores the binary data value in the form of electrical charge on a capacitor~\yct{(\vdd{} or 0~V)} and this data is accessed through an access transistor, driven by the wordline to \omcr{1}{connect} the cell capacitor to the bitline.

\head{DRAM Operation}
Data stored in a subarray is accessed \omcr{1}{at} row granularity.
 To access a row, the memory controller issues an \act{} \omcr{1}{(\texttt{ACTIVATE})} command to assert the wordline and enable the sense amplifier. When the wordline is asserted, the cell capacitor connects to the bitline and shares its charge, causing a \ieycr{1}{perturbation} on the bitline voltage. After the sense amplifier is enabled, it senses and amplifies the voltage deviation \ieycr{0}{on the bitline} towards \vdd{} or \gfcr{1}{0~V}. Once the data is fetched to the sense amplifiers, the memory controller may issue \wri{}/\rd{} commands to write to/read from the row. To access another row, the bank needs to be in the \omcr{1}{precharged} state. To do so, the memory controller issues a \pre{} \omcr{1}{(\texttt{PRECHARGE})} command to disable \gfcr{1}{the} sense amplifiers, de-assert the wordline, and precharging the bitlines \ieycr{1}{to \vddh{}}. Once the bank is precharged, the memory controller can access another row.

\head{DRAM Timing}
To ensure correct operation, the memory controller must obey the DRAM timing parameters specified in the DRAM interface standards (e.g., DDR4) by \gf{the} \gls{jedec}~\cite{jedec2017ddr4}.
We describe the most relevant timing constraints in the scope of this paper. The memory controller must wait for \gls{tras} before issuing a \pre{} command after an \act{} command. To open another row, the memory controller must wait for \gls{trp} before issuing another \act{} command. \omcr{1}{For more details on DRAM timing, we refer the reader to prior works~\cite{lee2015adaptive,kim2012case,chang2016lisa,chang2017understanding}.}

\subsection{\ieycr{0}{Processing Using COTS DRAM Chips}}
\label{sec:back_pud}
\head{Simultaneous Many-Row Activation} 
\omcr{1}{Current} DRAM \omcr{1}{standards} do \emph{not} officially support \gls{pud} operations\omcr{1}{. Yet,} the design of \ieycr{0}{COTS} DRAM chips does \emph{not} prevent users from activating multiple \gf{DRAM rows} at once by issuing an \apaLong{} command sequence \omcr{1}{(called \apa{})} with violated \gls{tras} and \gls{trp} timing constraints~\cite{gao2019computedram, olgun2021quactrng, gao2022frac, yaglikci2022hira, olgun2021pidram,olgun2023dram}. By doing so, two fundamental \gls{pud} operations can be performed in \ieycr{0}{COTS} DRAM chips: \one{} \gf{in-DRAM majority-of-three} \omcr{1}{(as in Ambit~\cite{seshadri2017ambit,seshadri2015fast})} and \two{} \gf{in-DRAM row copy (\omcr{1}{as in} RowClone~\cite{seshadri2013rowclone})}.

\noindent
\textbf{\gf{In-DRAM Majority-of-Three \ieycr{1}{(\maj{3})}}.} 
Prior \omcr{1}{works}~\cite{seshadri2017ambit,seshadri2015fast} introduces the concept of simultaneously activating three rows in a DRAM subarray (i.e., triple-row activation) \ieycr{1}{through modifications to the DRAM circuitry. When three rows are concurrently activated,} three cells connected to each bitline share charge simultaneously and contribute to the perturbation of the bitline~\cite{seshadri2017ambit,seshadri2015fast}. Upon sensing the perturbation of the three simultaneously activated rows, the sense amplifier amplifies the bitline voltage to $V_{DD}$ or 0~V if at least two of the three DRAM cells are charged or discharged, respectively. As such, simultaneously activating three rows results in a Boolean majority-of-three operation (\maj{3}). \ieycr{1}{Prior works~\cite{gao2019computedram,olgun2023dram,gao2022frac} demonstrate that COTS DRAM chips are capable of performing \maj{3} by simultaneously activating multiple rows in the same subarray. The state-of-the-art mechanism, FracDRAM~\cite{gao2022frac}, shows that a DRAM cell in COTS DDR3 chips can store fractional values (e.g., \vddh{}). FracDRAM uses fractional values to perform \maj{3} operations while simultaneously activating four DRAM rows in the same subarray.}

\noindent
\textbf{\gf{In-DRAM Row Copy}.} RowClone~\cite{seshadri2013rowclone} enables data movement within DRAM \omcr{1}{at} row granularity without incurring the energy and execution time costs of transferring data between the DRAM and the computing units. \omcr{1}{An intra-subarray} RowClone operation~\cite{seshadri2013rowclone} works \gfcr{1}{by} issuing \gfcr{1}{two} back-to-back \act{} commands: the first \act{} copies the contents of the source row $A$ into the \gfcr{1}{sense amplifiers}. The second \act{} connects the DRAM cells in the destination row~$B$ to the bitlines. Because the sense amplifiers have already sensed and amplified the source data by the time row~$B$ is activated, the data in each cell of row~$B$ is overwritten by the data stored in the \ieycr{0}{sense amplifiers} (i.e., row~$A$'s data). \omcr{1}{Building on this observation,} prior works~\cite{gao2019computedram,olgun2021pidram} \ieycr{0}{experimentally} demonstrate \omcr{1}{that} \omcr{1}{COTS DRAM chips are capable of performing} \gf{RowClone operation by enabling} consecutive activation of two DRAM rows \ieycr{1}{in the same subarray}.

\setcounter{version}{1}
\section{Methodology}
\label{sec:method}

\ieycr{0}{We describe our methodology for two analyses. First, we experimentally characterize the computational capability of \nCHIPS{} commercial off-the-shelf (COTS) DRAM chips from two major manufacturers in terms of simultaneously activating many DRAM rows (\secref{sec:mra_method}), performing majority operations (\secref{sec:maj_method}), and copying one row's content (concurrently) to multiple rows, i.e., \mrc{} (\secref{sec:mri_method}). Second, to back up our observations from real-device experiments, we conduct SPICE~\cite{ltspice,vladimirescu1994spice} simulations (\secref{sec:spice_method}).}

\subsection{DRAM Testing Infrastructure}
\label{sec:dram_testing_infra}
\begin{figure}[ht]
\centering
\includegraphics[width=1\linewidth]{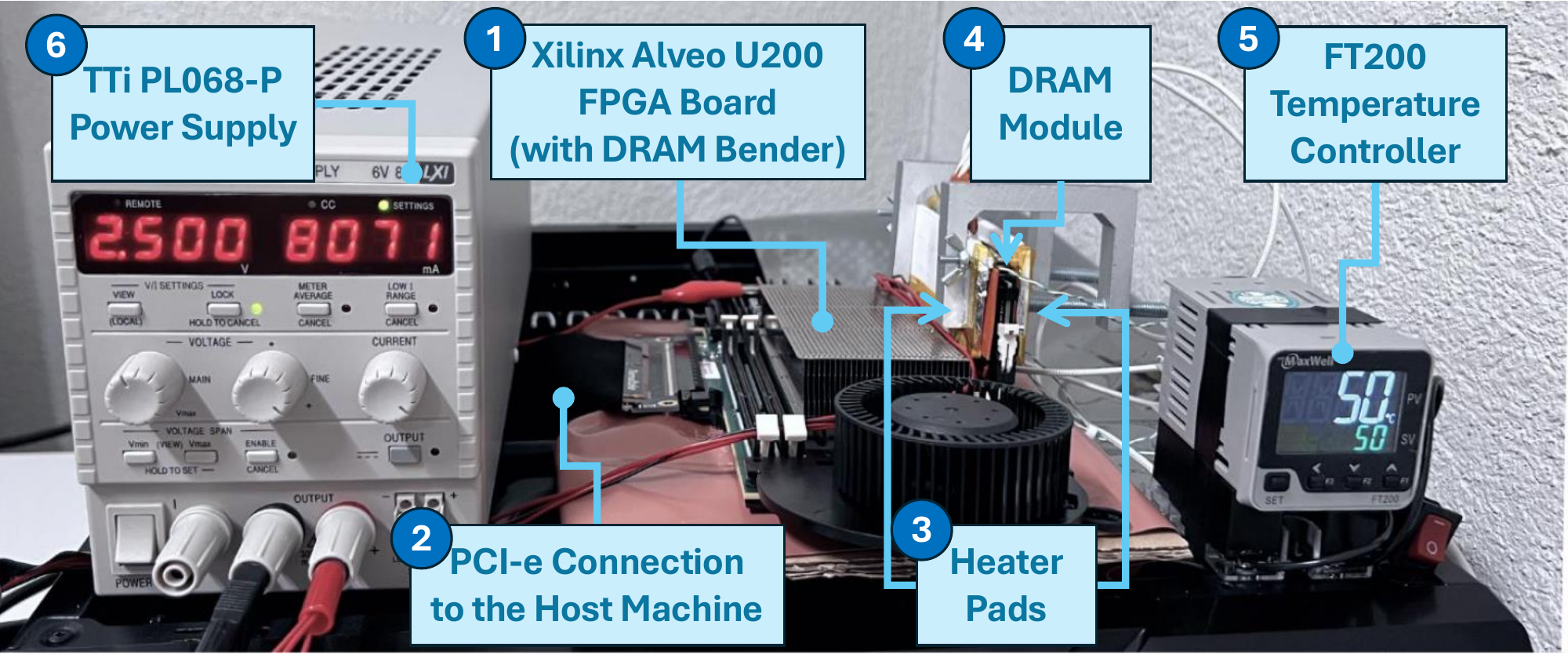}
\caption{DDR4 DRAM Bender~\cite{olgun2023dram} experimental setup.}
\label{fig:infra}
\end{figure}

We conduct real DRAM chip experiments \gf{using} DRAM Bender~\cite{safari-drambender, olgun2023dram}, a DDR4 testing infrastructure that provides precise control of the DRAM commands issued to a module. \figref{fig:infra} shows our experimental setup\gf{, which} consists of \ieycr{1} {six} main components: \textcolor{blue}{\dingOne{}} \omcr{1}{a} Xilinx Alveo U200 FPGA board~\cite{alveo_u200} programmed with DRAM Bender, \textcolor{blue}{\dingTwo{}} a host machine that generates DRAM commands \nb{used} in our tests, \textcolor{blue}{\dingThree{}} rubber heaters that clamp the \textcolor{blue}{\dingFour{}} DRAM module on both sides to avoid fluctuations in ambient temperature, \textcolor{blue}{\dingFive{}} a MaxWell FT200 temperature controller~\cite{maxwellFT200} that keeps the DRAM chips at the target temperature, and \textcolor{blue}{\dingSix{}} an external TTi PL068-P power supply~\cite{pl068p}, which enables us to control \gf{DRAM wordline voltage (i.e., }\vpp{}\gf{)} at the precision of ±1 mV.\footnote{\gf{Modern DRAM chips use two separate voltage rails~\cite{jedec2017ddr4,jedec2020ddr5,jedec2016gddr5x,jedec2021gddr6,yaglikci2022understanding}: {1)~\vdd, used to operate the core DRAM array and peripheral circuitry, and 
2)~\vpp{}, used exclusively to assert a wordline during a DRAM row activation.}}}

\head{Real DDR4 DRAM Chips Tested}
Table~\ref{tab:dram_chips} shows the \param{\nCHIPS{}} (\param{\nMODULES{}}) real DDR4 DRAM chips (modules) \ieycr{0}{from two major DRAM manufacturers that we focus our analysis on.} To demonstrate that our observations are not specific to a certain DRAM architecture/process but rather common across different designs and generations, we test a variety of DRAM
chips spanning different die densities and die revisions.\ieycr{0}{\footnote{We also test Samsung chips; however, we observe no successful \gls{pud} operations. \ieycr{1}{We hypothesize why we do not observe all tested \gls{pud} operations in Samsung chips in~\secref{sec:limitations}.} We provide much more detail on all tested DRAM chips in the extended version of this paper~\cite{yuksel2024simultaneous}.}} 
\begin{table}[ht]
\centering
\caption{Summary of DDR4 DRAM chips tested.}
\label{tab:dram_chips}
\resizebox{\linewidth}{!}{%
\begin{tabular}{ccccccc}
\textbf{DRAM Mfr.} & \textbf{\#Modules} & \textbf{\#Chips} &  \textbf{Die Rev.} & \textbf{Density} & \textbf{Org.} & \textbf{\ieycr{0}{Subarray} Size}  \\
\hline\hline
SK Hynix & 7 & \param{56} & M & 4Gb  & x8  & 512 or 640 \\
(Mfr. H) & 5 & \param{40} & A & 4Gb  & x8  & 512 \\ 
\midrule
Micron   & 4 & \param{16} & E  & 16Gb & x16 & 1024 \\
(Mfr. M) & 2 & \param{8} & B  & 16Gb & x16 & 1024  \\
\bottomrule
\end{tabular}
}
\end{table}

\head{Metric} We define a metric to evaluate \nb{the robustness of} \gls{pud} operations:~\emph{the success rate}. Success rate refers to the percentage of DRAM cells that produce correct output in \ieycr{1}{\emph{all}} \atbcr{1}{test} trials \nb{of a \gls{pud} operation. Hence, we have two data points here.} If a \gf{DRAM} cell produces an incorrect result at least once, we refer to this \gf{DRAM} cell as an \emph{unstable cell} that \emph{cannot} be used to perform \gls{pud} operations. For example, if an operation has a 25\% success rate, it means \ieycr{1}{that} a quarter of the \gf{DRAM} cells \nb{always} produce correct results in the simultaneously activated rows and can be used to \nb{reliably} perform that operation. We report the success rate distribution, which comes from all tested row groups in all DRAM chips.

\head{\pmb{\gls{tras}} and \pmb{\gls{trp}} Scaling} We test various reduced timing delays \one{} between \act{} and \pre{} commands, i.e., $t_{1}$, and \two{} between \pre{} and \act{} commands, i.e., $t_{2}$. All experiments are conducted at the timing delays that achieve the highest \ieycr{2}{success rate} for the tested \gls{pud} operations unless stated otherwise.

\head{Temperature Scaling} We perform our experiments \omcr{1}{at} five temperature levels: 50$^{\circ}$C, 60$^{\circ}$C, 70$^{\circ}$C, 80$^{\circ}$C, and 90$^{\circ}$C. All experiments are conducted at 50$^{\circ}$C \nb{unless stated otherwise}.

\head{$\pmb{V_{PP}}$ Voltage Scaling} We test five \vpp{} levels: 2.5V, 2.4V, 2.3V, 2.2V, and 2.1V.
All experiments are conducted at the nominal \vpp{} level (i.e., 2.5V) \nb{unless stated otherwise}. 
\gf{In our analysis, we focus on the effects of \vpp{} on \gls{pud} computation. Even though both \vpp{} and \vdd{} can affect a DRAM chip’s computation robustness, changing \vdd{} can negatively impact DRAM reliability in ways \omcr{1}{unrelated} to computation \atbcr{1}{inside a DRAM subarray} (e.g., I/O circuitry instabilities) because \vdd{}
supplies power to all logic elements within the DRAM chip. In contrast, \vpp{}
affects only the wordline voltage. \atbcr{1}{Therefore, we expect} \vpp{} \atbcr{1}{to} influence computation \atbcr{1}{inside a subarray}, without adversely affecting \atbcr{1}{DRAM chip components outside a subarray}.~\atbcr{1}{\vpp{} also affects data retention~\cite{yaglikci2022understanding} and access latency characteristics~\cite{chang_understanding2017} of a DRAM chip. Such effects of \vpp{} are out of the scope of our analyses.}}

\noindent
\textbf{Data Pattern\ieycr{0}{s}.}~\ieycr{0}{We use two types of data patterns: random and fixed data patterns. For the random data pattern, we generate} uniformly distributed random data and \ieycr{0}{fill each activated row with the random data, where each activated row has a different data pattern. For fixed data patterns} (i.e., 0x00/0xFF, 0xAA/0x55, 0xCC/0x33, 0x66/0x99), \ieycr{0}{we fill each activated row either with} 1) all 0x00 or all 0xFF, 2) all 0xAA or all 0x55, 3) all 0xCC or all 0x33, and 4) all 0x66 or all 0x99. All experiments are conducted using the random data pattern \ieycr{1}{(where we observe the lowest success rate across tested data patterns)} unless stated otherwise.

\head{Number of Instances Tested}
\ieycr{1}{We randomly select three subarrays in each bank (a total of 16 banks) per DRAM module. Within each subarray, we randomly test 100 different groups of rows that are simultaneously activated each for 2-, 4-, 8-, 16-, and 32-row activation, which results in testing a total of 24K different groups of simultaneously activated rows per module.}

\head{Finding Subarray Boundaries}
\iey{To understand the computational capability of \gls{pud} operations performed in a \gf{DRAM} subarray,} \nb{we follow the methodology used in prior works to \omcr{1}{identify} subarray boundaries~\cite{yaglikci2022hira, gao2019computedram, gao2022frac, olgun2021quactrng, olgun2021pidram,olgun2023dram}. We leverage the observation that it is possible to copy a row's data to another row (i.e., RowClone operation~\cite{seshadri2013rowclone}) within the same subarray by leveraging the shared {bitlines}. We repeatedly perform RowClone \omcr{1}{across} all row \omcr{1}{pairs}. If we can copy a row's data to another row, we infer that these two rows are within a subarray. By conducting this experiment, we reverse engineer the subarray sizes (listed in Table~\ref{tab:dram_chips}) and subarray boundaries in all tested DRAM modules.}

\subsection{Simultaneous Many-Row Activation Experiments}
\label{sec:mra_method}
\vspace{-2pt}
\ieycr{1}{In our experiments, we use the \apaTime{} (\apa{}) DRAM command sequence, where \rfirst{} is the firstly activated row (i.e., $Row_{First}$), \rsecond{} is the secondly activated row (i.e., $Row_{Second}$), and $t_{1}$ and $t_{2}$ are timing delays between command pairs in the command sequence.}

\head{Testing Methodology} 
\ieycr{1}{Our experiment consists of three steps. First, we initialize a whole subarray with a predefined data pattern. Second, we issue \atbcr{1}{an \apa{}} command sequence to simultaneously activate multiple rows in the subarray. Third, we issue a \wri{} command with a different data pattern from the predefined data pattern by respecting the timing parameters. This \wri{} command causes the sense amplifiers to overdrive their bitlines and thus updates the values of the cells in all simultaneously activated DRAM rows~\cite{olgun2021quactrng,dram-circuit-design}}. After the three-step procedure, we precharge the bank and read each row in the subarray while adhering to the nominal timing parameters. When we read all rows in the subarray, we expect that simultaneously activated rows store the exact data pattern sent with the \wri{} command. 

\ieycr{1}{We test every possible \rfirst{} and \rsecond{} combination in the \apa{} command sequence and various $t_{1}$ and $t_{2}$ timing parameters for each tested subarray. We observe that not all simultaneously activated rows fully store the exact data pattern (i.e., simultaneous many-row activation is not 100\% robust.). Hence, we report the success rate of this simultaneous many-row activation as the percentage of simultaneously activated rows' cells that store the exact data pattern sent with the \wri{} command.}

\subsection{\omcr{1}{Bitwise} Majority \omcr{1}{Computation} Experiments}
\label{sec:maj_method}

\head{\ieycr{1}{Key Idea}}~\ieycr{0}{\ieycr{1}{Our key idea to} perform \omcr{1}{an} in-DRAM majority-of-X (\maj{X}) operation \ieycr{1}{consists of} four steps.}
First, we issue an \act{}~\rfirst{} command to assert the wordline of the \rfirst{}, \nb{connecting \rfirst{}} to the bitline. 
Second, we issue \atbcr{1}{a} \pre{} command immediately after the first \act{} \omcr{1}{while} \nb{greatly violating \gls{tras}}. This way, \rfirst{} does not have sufficient time to \atbcr{1}{fully} share its charge with \nb{the} bitline. Third, we issue the second \act{} command to another row by greatly violating \gls{trp}. This prevents \omcr{1}{the DRAM control circuitry from} de-asserting \rfirst{} and activates multiple rows. \nb{All activated rows share their charge with the bitline, \omcr{1}{which} result\omcr{1}{s} in a \omcr{1}{perturbation in the bitline that corresponds to a} majority operation across these rows.} Finally, the sense amplifier amplifies the \omcr{1}{perturbation} on the bitline. If the \maj{X} is successful, the sense amplifier overwrites the bitline \nb{and all activated rows' content with} the \maj{X}'s result.~\ieycr{0}{For example, if we activate three \omcr{1}{cells} at once\ieycr{2}{,} and they have A, B, and B values, after a successful \maj{3} operation, these three \omcr{1}{cells} \nb{would} have B \nb{as their} value.}

\head{Replicating Inputs of \maj{X}}
\gf{\ieycr{1}{To perform \maj{X} operation with N-row activation (i.e., simultaneously activating N rows)}, we replicate \atbcr{1}{each} \maj{X} input \ieycr{1}{operand (i.e., a total of X input operands) $\lfloor N/X \rfloor$ times}.\footnote{{In Boolean algebra, when \maj{} inputs are replicated, \maj{} maintains the functionality, e.g., \maj{6(A,B,C,A,B,C)}=\maj{3(A,B,C)}.}} For example, if we perform \maj{3} with 32-row activation, we replicate each input ten \ieycr{1}{(i.e., $\lfloor 32/3 \rfloor = 10$)} times. If the number of DRAM rows opened is not a multiple of X \ieycr{1}{(i.e., $N\%X > 0$)}, we \atbcr{1}{initialize $N\%X$ of the activated DRAM rows in a way that they do \emph{not} contribute to bitline voltage,} 
\ieycr{1}{using \atbcr{1}{a} \texttt{Frac}~\cite{gao2022frac} operation} (\secref{sec:back_pud}).\footnote{We use \texttt{Frac} parameters that achieve the best success rate for \maj{X} operation (i.e., the number of \texttt{Frac} operations, the timing delay, and the location of neutral rows). We refer the reader to FracDRAM~\cite{gao2022frac} for more detail.} \atbcr{1}{We call these rows \emph{neutral rows}}. For example, if we perform \maj{3} with 32-row activation, \ieycr{1}{we set the remaining two ($32\%3 = 2$) DRAM rows neutral}.}

\head{Testing Methodology} 
We characterize \omcr{1}{the} \maj{X} operation in \nb{five} steps: \one{} \ieycr{1}{store \maj{X}'s input operands (a total of X operands) in X rows}, \two{} replicat\omcr{1}{e} inputs of \maj{X} into remaining simultaneously activated rows \three{} \atbcr{1}{initialize neutral rows using \ieycr{2}{the} \texttt{Frac} operation, if it is needed,} \four{} perform \omcr{1}{the} \maj{X} operation \ieycr{1}{by issuing \apaTime{} command sequence}, \five{} wait for \gls{trp}, and \six{} read back the values in the row buffer.\footnote{For Mfr. M, \texttt{Frac} operation is not supported. However, we observe that the sense amplifiers of these modules are always biased to one or zero. Initializing the neutral rows with all zeros/ones enables \omcr{1}{a \maj{X}} operation.}

\subsection{Multi-RowCopy Experiments}
\label{sec:mri_method}

\head{\ieycr{1}{Key Idea}} 
\nb{Copying one source DRAM row's content to multiple destinations DRAM rows at once (\emph{\mrc{}})} \omcr{1}{comprises} of \nb{four} steps \omcr{1}{in our real chip evaluation}. 
First, we issue \act{} to \nb{the source row (i.e., \rfirst{})} to assert the wordline and to connect \nb{\rfirst{}} to the bitline\omcr{1}{s}.
Second, we issue \pre{} after waiting for \gls{tras}. This ensures the sense amplifier senses \rfirst{} correctly and drives bitlines to the source row's charge.
Third, we issue the second \act{} command by \ieycr{0}{greatly} violating \gls{trp} \ieycr{1}{(i.e., $\leq 3ns$).}\footnote{\ieycr{1}{Prior works~\cite{gao2019computedram,olgun2021pidram,olgun2023dram} perform in-DRAM copy operation from one row to another using \apa{} by waiting for more between \pre{} and \act{} than \mrc{} does (e.g., for 6ns). This results in consecutive activation of two rows.}} The second \act{} command interrupts the \pre{} command. By doing so, it \one{} prevents the bitline from being precharged to \vddh{}, 2) keeps \nb{\rfirst{}} and the sense amplifier enabled, and 3) simultaneously \ieycr{0}{activates} many rows. Finally, the sense amplifier overwrites all activated rows with \nb{the source row's} data. If the \mrc{} is successful, all the simultaneously activated rows with an \apa{} command have the source row's data.

\head{Testing Methodology} We characterize \omcr{1}{the} \mrc{} operation in \param{four} steps: we 1) initialize 
\atbcr{1}{all destination rows} with a predetermined data pattern, 2) initialize the source row with a different data pattern \ieycr{1}{from} the predetermined data pattern, 3) perform \ieycr{1}{the} \mrc{} operation, 4) precharge the bank and individually read each \atbcr{1}{destination row} 
while adhering to the manufacturer-recommended DRAM timing parameters.

\subsection{SPICE Simulations}
\label{sec:spice_method}

To provide insights into our real-chip-based experimental observations, we conduct SPICE~\cite{vladimirescu1994spice} simulations to measure the bitline and cell voltage levels. We extend a DRAM model used in prior work~\cite{chang_understanding2017} that allows us to perform multiple-row activation. We use LTspice~\cite{ltspice} with the reference 55 nm DRAM model from Rambus~\cite{rambus_model} and scale it based on the ITRS roadmap~\cite{itrs_model,vogelsang2010understanding} to model the 22 nm technology node following the PTM transistor models~\cite{ptm2012transistors}.\footnote{We do \emph{not} expect SPICE simulation and real-world experimental results to be identical because a circuit model cannot simulate a real DRAM chip’s exact behavior without proprietary design and manufacturing information.} To account for process variation, we perform Monte-Carlo simulations~\cite{james1980monte} over $10^4$ iterations by randomly varying the capacitor and transistor parameters \param{10}\SI{}{\percent}, \param{20}\SI{}{\percent}, \param{30}\SI{}{\percent}, and \param{40}\SI{}{\percent} for each run.

\setcounter{version}{2}
\section{\ieycr{0}{Characterization of Simultaneous Many-Row Activation
\omcr{2}{in COTS DRAM Chips}}}
\label{sec:sim-act}
\ieycr{0}{We provide two key analyses on simultaneous many-row activation in COTS DRAM chips. First, w}e characterize \mel{the} \yct{robustness of} simultaneous many-row activation by analyzing the effect of 1)~timing delay between each command \omcr{1}{pair} in \omcr{1}{the} \gls{apa} \omcr{1}{sequence}, 2)~\ieycr{0}{DRAM chip temperature}, and 3)~wordline voltage underscaling. \ieycr{0}{Second, we analyze how the power consumption of simultaneous many-row activation measures against standard DRAM operations.}

\begin{figure}[ht]
\centering
\includegraphics[width=\linewidth]{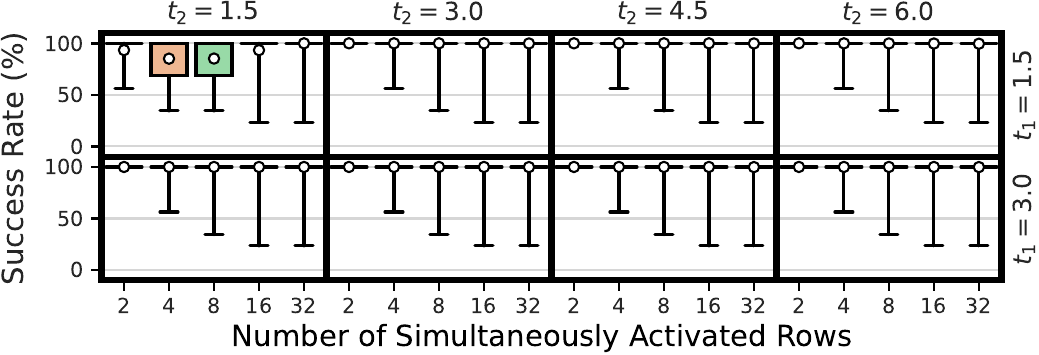}
\caption{\ieycr{0}{E}ffect of $t_{1} (ns)$ and $t_{2} (ns)$ on the success rate of simultaneous many-row activation.}
\label{fig:timing_sim_act}
\end{figure}

\label{sec:sim-act-char}
\head{Effect of Timing Delay}
\figref{fig:timing_sim_act} shows the distribution of the success rate for different numbers of \omcr{1}{simultaneously} activated rows across all tested row groups in all DRAM chips in a box and whiskers plot for different combinations of $t_{1}$, the timing between \act{} and \pre{} \yct{(rows of subplots)} and $t_{2}$, the timing delay between \pre{} and \act{} \yct{(columns of subplots)}.\footnote{\label{fn:boxplot}{In a box-and-whiskers plot, the box is lower-bounded by the first quartile and upper-bounded by the third quartile. The \gls{iqr} is the distance between the first and third quartiles (i.e., box size). Whiskers show the minimum and maximum values.}} \ieycr{0}{We make Observations 1 and 2 from \figref{fig:timing_sim_act}.}

\observation{\ieycr{0}{COTS} DRAM chips can simultaneously activate up to 32 rows \ieycr{1}{with a >99.85\% success rate}.}
When we issue \omcr{1}{an} \gls{apa} with the best timing delay (i.e., $t_{1}=3ns,\ t_{2}=3ns$), we can perform 2-, 4-, 8-, 16-, and 32-row activation with an average success rate of \param{99.99}\%, \param{99.99}\%, \param{99.99}\%, \param{99.99}\%, and \param{99.85}\%. We derive Takeaway 1 from Observation 1.
\takeaway{\ieycr{0}{COTS} DRAM chips are capable of simultaneously activating 2, 4, 8, 16, and 32 rows \ieycr{1}{at very high success rates}.}

\observation{If $t_{1}$ \ieycr{0}{or} $t_{2}$ are lower than 3ns, we observe a drastic decrease in success rate.}
For example, for 8-row activation, choosing $t_{1}$=$t_{2}$=1.5ns decreases the average success rate by \param{21.74}\% compared to the best timing delay configuration \ieycr{1}{(i.e, $t_{1}$=1.5ns and $t_{2}$=3ns)}. We hypothesize that the row decoder \ieycr{0}{circuitry}~\cite{dram-circuit-design} \ieycr{0}{cannot fully activate all cells in the to-be-activated} rows due to the very low delay ($t_{1}$+$t_{2}$=3ns) between two \act{}s \ieycr{0}{in \apa{} command sequence}.

\begin{figure}[ht]
\centering
\begin{subfigure}[b]{0.5\linewidth}
\centering
\includegraphics[width=\linewidth]{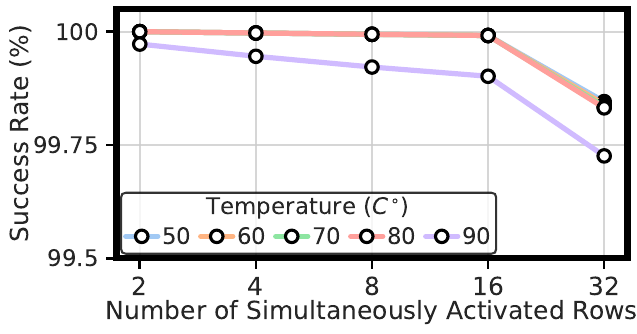}
\caption{}
\label{subfig:temperature_sim_act}
\end{subfigure}
\begin{subfigure}[b]{0.5\linewidth}
\centering
\includegraphics[width=\linewidth]{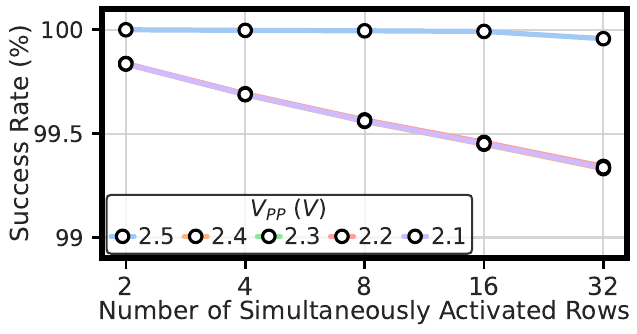}
\caption{}
\label{subfig:voltage_sim_act}
\end{subfigure}
\caption{\ieycr{0}{Average success rate} of simultaneous many-row activation with \omcr{1}{varying} (a) temperature and (b) wordline voltage.}
\label{fig:temperature_voltage_sim_act}
\end{figure}

\head{Temperature Scaling}
\figref{subfig:temperature_sim_act} shows the average success rate (y-axis) of simultaneously activating various numbers of rows (x-axis) under five temperature levels: \SI{50}{\celsius}, \SI{60}{\celsius}, \SI{70}{\celsius}, \SI{80}{\celsius}, and \SI{90}{\celsius}. \ieycr{0}{We make Observation 3 from \figref{subfig:temperature_sim_act}.}

\observation{Increasing temperature up to \SI{90}{\celsius} has a \ieycr{0}{small} effect on the success rate.}
We observe only \omcr{1}{a} \param{0.07}\% \omcr{1}{average} success rate decrease on average on simultaneous many-row activation when the temperature is increased from \SI{50}{\celsius} to \SI{90}{\celsius}. We hypothesize that since simultaneous many-row activation experiments use I/O peripheral circuitry to perform \wri{} commands, temperature could affect not only the subarray elements and \omcr{1}{their} operation (e.g., DRAM cell, sense amplifiers, and charge-sharing) but also peripheral circuitry and \omcr{1}{their} operation (e.g., write drivers/buffers and driving bitlines with the data pattern sent with \wri{} command~\cite{dram-circuit-design}). We believe that fully understanding the effect of temperature on the reliability of simultaneous many-row activation begs more investigation at the circuit level, which we leave for future work.

\head{Voltage Scaling}
\figref{subfig:voltage_sim_act} shows the average success rate (y-axis) of simultaneously activating various numbers of rows (x-axis) under five \vpp{} levels (in different line colors): 2.5V, 2.4V, 2.3V, 2.2V, and 2.1V.\footnote{\label{fn:voltage_fn}We test two DRAM modules due to time and infrastructure limitations.} \ieycr{0}{We make Observation 4 from \figref{subfig:voltage_sim_act}.}

\observation{\omcr{1}{Reducing} the wordline voltage \omcr{1}{only} slightly affects the success rate of simultaneous many-row activation.}
 Underscaling voltage from 2.5V to 2.1V decreases the average success rate by at most \param{0.41}\%.

\ieycr{0}{We derive Takeaway 2 from Observations 3 and 4.}
\takeaway{\omcr{1}{S}imultaneous many-row activation is highly resilient to temperature and wordline voltage \omcr{1}{in COTS DRAM chips}.}

\begin{figure}[ht]
\centering
\includegraphics[width=0.6\linewidth]{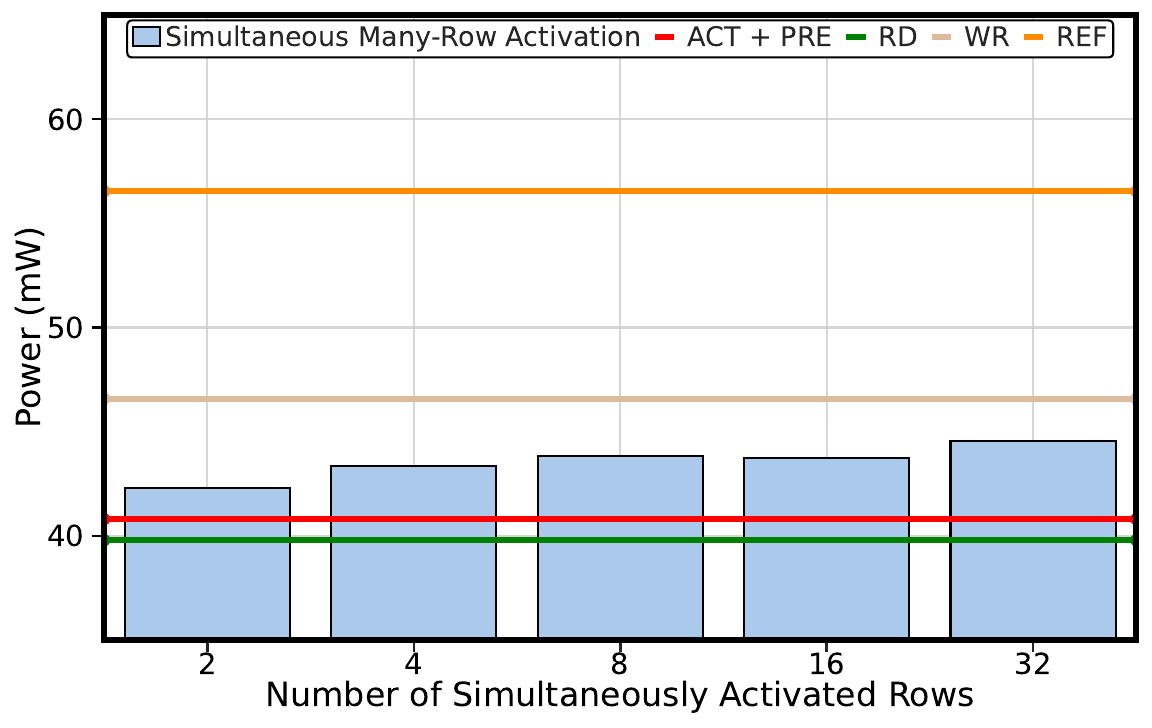}
\caption{\ieycr{2}{Power consumption of simultaneous many-row activation and standard DRAM operations.}}
\label{fig:power_sim_act}
\end{figure}

 \head{Power Consumption} We evaluate the power consumption of simultaneous many-row activation operations and standard DRAM operations (i.e., \rd{}, \wri{}, \act{}+\pre{}, and \texttt{REF}) using one DRAM module.\footnote{\ieycr{2}{We provide a detailed description of how we conduct this experiment in the extended version of this paper~\cite{yuksel2024simultaneous}}}~\figref{fig:power_sim_act} shows the average power consumption measured for each operation. Dashed lines represent standard DRAM operations. \ieycr{0}{We make Observation 5 from \figref{fig:power_sim_act}.}

\observation{Simultaneous many-row activation power draw likely meets the power budget of DDR4 chips.}
The power consumption of simultaneously activating 32 rows is \param{21.19}\% smaller than the most power-consuming single DRAM operation's (i.e., \texttt{REF}) power consumption.

\setcounter{version}{2}
\section{\ieycr{0}{Characterization of} Majority Operations \omcr{1}{in COTS DRAM Chips}}
We experimentally characterize the robustness of \ieycr{0}{majority-of-X} (\maj{X}) operations, where $X \in\{3,5,7,9\}$, under four operating parameters: \one{} timing delay between \omcr{1}{each} command \omcr{1}{pair} in the \apa{} sequence, \two{} data pattern, \three{} \ieycr{0}{DRAM chip} temperature, and \four{} wordline voltage.

\label{sec:maj_char}

\begin{figure}[ht]
\centering
\includegraphics[width=\linewidth]{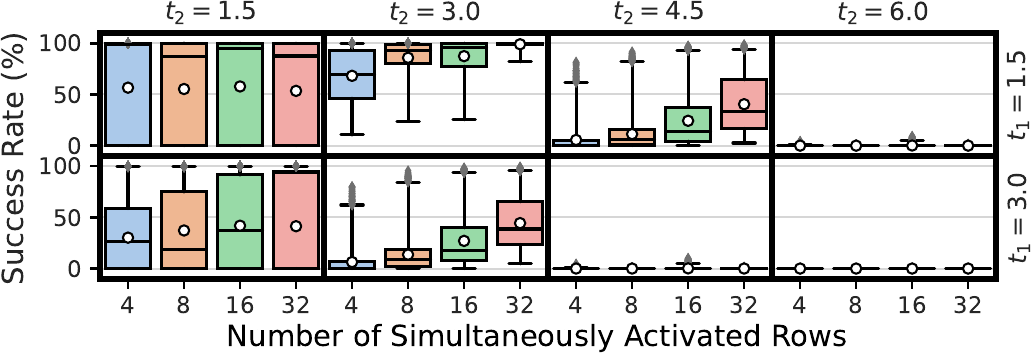}
\caption{Effect of \ieycr{0}{$t_{1}(ns)$, $t_{2}(ns)$, and the number of simultaneously activated rows} on \maj{3} operation.}
\label{fig:timing_maj}
\end{figure}

\head{Effect of Timing Delay}
\figref{fig:timing_maj} shows the distribution of the \maj{3}'s success rate for different numbers of \ieycr{0}{simultaneously} activated rows in a box and whiskers plot for different combinations of $t_{1}$ (i.e., the timing between \act{} and \pre{}) and $t_{2}$ \nb{(i.e.,} the timing delay between \pre{} and \act{}).\footref{fn:boxplot} \ieycr{1}{We replicate the \maj{3}'s input operands 1, 2, 5, and 10 times when performing the \maj{3} operation with 4-, 8-, 16-, and 32-row activation, respectively (\secref{sec:maj_method})}. \ieycr{0}{We make Observations 6 and 7 from \figref{fig:timing_maj}.}

\observation{Storing multiple copies of \maj{3}'s input \ieycr{0}{operands} drastically increases the success rate \ieycr{0}{of \maj{3}.}}
\maj{3} with 32-row activation achieves \param{30.81}\% higher success rate than \maj{3} with 4-row activation. We hypothesize that \ieycr{1}{\maj{3} with 4-row activation has \atbcr{1}{a lower} success rate than \maj{3} with 32-row activation because \atbcr{1}{the deviation in a bitline's voltage is less likely to exceed the reliable sensing margin.}
\atbcr{1}{For such a bitline,}
the sense amplifier fails to reliably produce correct \maj{X} results.
Storing multiple copies of \maj{3}'s input \ieycr{0}{operands} could increase the deviation on the bitline voltage,
which results in an increased success rate} (the circuit-level simulations in \secref{sec:maj_hypo} \atbcr{1}{supports our} observation \ieycr{0}{and hypothesis}). 
\observation{Timing delays \omcr{1}{between commands in} the \apa{} command sequence heavily affects the success rate of \maj{3}.}
\nb{C}hoosing $t_{1}=1.5ns$ and $t_{2}=3ns$ to perform \maj{3} with 32-row activation achieves \param{99.00}\% average success rate, which is \param{45.50}\% higher than the second highest average success rate achieved by a different timing configuration \ieycr{0}{(i.e., $t_{1}=t_{2}=3ns$).} 
We have two hypotheses to explain Observation \param{7}. First, every activated row \ieycr{1}{should} contribute equally to the charge-sharing process to achieve the highest possible success rate in \ieycr{2}{the} \maj{X} operation. To achieve this, the timing delay between two \act{} commands ($t_1$ + $t_2$) should be kept as \omcr{1}{low} as possible since increasing this delay could cause the first activated row of the \apa{} sequence (i.e., $R_F$) to share its charge more than others. Second, \omcr{1}{too small} \omcr{2}{a} delay between \pre{} and \act{} may prevent (e.g., choosing $t_2=1.5ns$) \omcr{1}{the} asserti\omcr{1}{on of} intermediate signals in the row decoder circuitry, leading to \omcr{1}{an inability to} simultaneously activate multiple rows. Consequently, choosing $t_1=1.5ns$ and $t_2=3ns$ achieves the highest success rate.

\begin{figure}[ht]
\centering
\includegraphics[width=\linewidth]{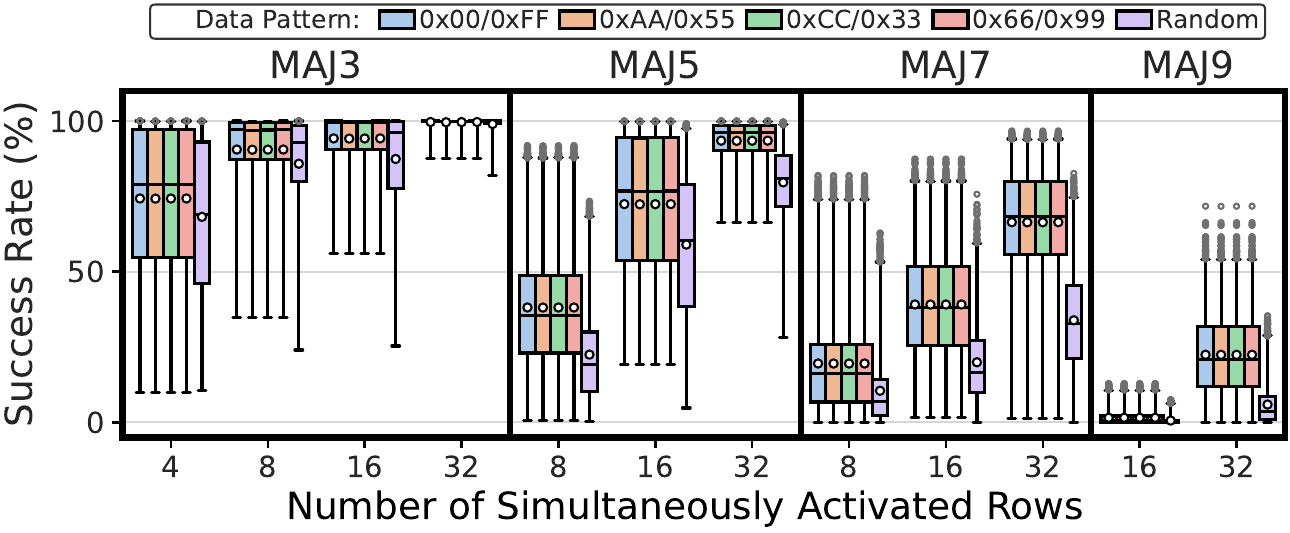}
\caption{\ieycr{0}{Success rates of \maj{3}, \maj{5}, \maj{7}, and \maj{9} operations with different data patterns.}}
\label{fig:data_maj}
\end{figure}

\head{Data Pattern} \ieycr{0}{We analyze the effect of data pattern on the success rate for  \maj{3}, \maj{5}, \maj{7}, and \maj{9} operations\footnote{We omit \maj{X} operations that have $<$1\% success rate at most (i.e., \maj{11+} for Mfr. H, and \maj{9+} for Mfr. M).} as we vary the number of simultaneously activated rows.} \figref{fig:data_maj} shows the success rate distribution of  \maj{3}, \maj{5}, \maj{7}, and \maj{9} operations \ieycr{0}{across DRAM cells for five tested} data patterns. \ieycr{2}{We make Observations 8-10 from \figref{fig:data_maj}.}

\observation{\ieycr{0}{We can perform \maj{5}, \maj{7}, and \maj{9} operations in COTS DRAM chips.}}
\ieycr{0}{COTS DRAM chips are capable of performing \maj{5}, \maj{7}, and \maj{9} operations. We observe that performing \maj{5}, \maj{7}, and \maj{9} operations with 32-row activation achieves average success rates \ieycr{2}{(across tested groups of rows that are simultaneously activated)} of \param{79.64}\%, \param{33.87}\%, and \param{5.91}\%, respectively.}

\ieycr{0}{We derive Takeaway 3 from Observation 8.}

\takeaway{\ieycr{0}{COTS} DRAM chips \ieycr{0}{are capable of} perform\ieycr{0}{ing} \maj{5}, \maj{7}, and \maj{9} operations.}

\observation{Data pattern significantly affects the success rate of \ieycr{2}{the} \maj{X} operation.}
\ieycr{0}{Fixed data patterns (i.e., 0x00/0xFF, 0xAA/0x55, 0xCC/0x33, and  0x66/0x99 data patterns) have a small and similar effect on the success rate of \ieycr{2}{the} \maj{X} operation, whereas uniformly distributed random data pattern significantly decreases the success rate of the \maj{X} operation.} \ieycr{0}{For example, when using 32-row activation, \maj{3}, \maj{5}, \maj{7}, and \maj{9} operations with random data pattern have \param{0.68}\%, \param{13.85}\%, \param{32.56}\%, and \param{16.51}\% lower average success rate\omcr{2}{s} than \maj{3}, \maj{5}, \maj{7}, and \maj{9} operations with all 0x00/0xFF data pattern, respectively.}

\observation{Storing multiple copies of \maj{X}'s \ieycr{0}{input operands} increases the success rate of not only \maj{3} but also \maj{5}, \maj{7}, and \maj{9} operations.}

\ieycr{2}{For example, s}toring multiple copies increases the average success rate of \maj{5}, \maj{7}, and \maj{9} \ieycr{2}{with random data pattern} by \param{56.27}\%, \param{35.15}\%, and \param{13.11}\%. 
 
\ieycr{0}{We derive Takeaway 4 from Observations 6 and 10.}
\takeaway{Storing multiple copies of \maj{X}'s input operands significantly increases the success rate of the \maj{X} operation \ieycr{0}{in COTS DRAM chips}.}

\begin{figure}[ht]
\centering
\includegraphics[width=\linewidth]{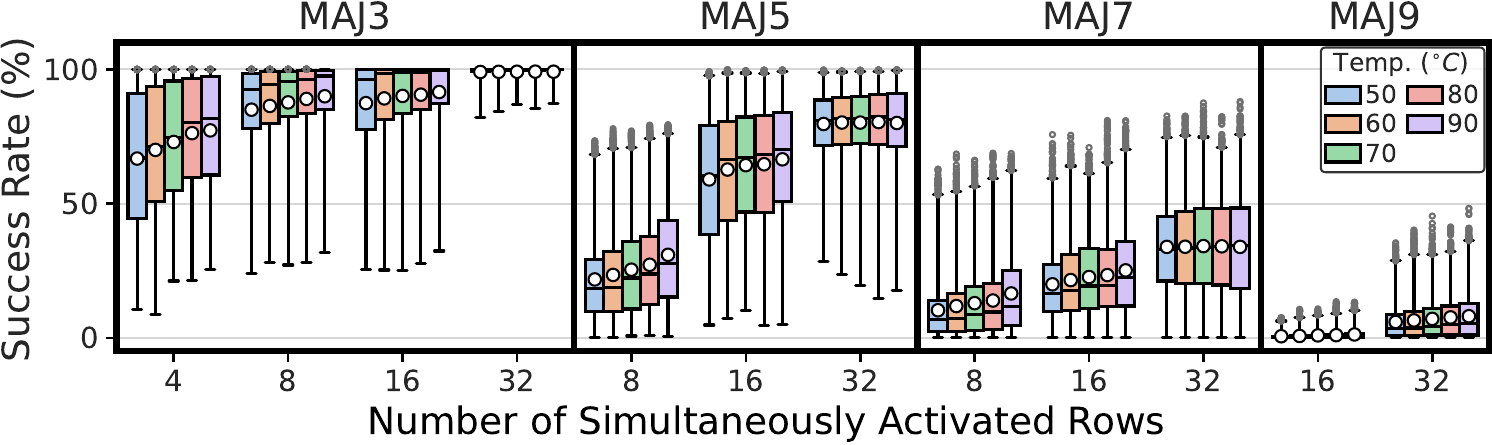}
\caption{\ieycr{0}{S}uccess rate of \maj{X} \ieycr{0}{at} \ieycr{0}{different DRAM chip} temperatures.}
\label{fig:temp_maj}
\end{figure}

\head{Temperature Scaling}
\figref{fig:temp_maj} shows the success rate distribution of \ieycr{2}{the} \maj{X} operation with various simultaneously activated rows \ieycr{0}{at} five \ieycr{0}{different} temperature levels: \SI{50}{\celsius}, \SI{60}{\celsius}, \SI{70}{\celsius}, \SI{80}{\celsius}, and \SI{90}{\celsius}. \ieycr{0}{We make Observations 11 and 12 from \figref{fig:temp_maj}.}

\observation{Temperature \param{slightly} affects the success rate \ieycr{0}{of} \omcr{2}{the} \maj{X} operation.}
For example, from \SI{50}{\celsius} to \SI{90}{\celsius}, the average success rate varies by \param{4.25}\% on average across all the tested operations.
We hypothesize that increasing the temperature could reduce the threshold voltage\omcr{2}{s} of cells' access transistor\omcr{2}{s}, and thus, the charge-sharing operation among activated rows and bitlines happens faster and stronger~\cite{sakata1994subthreshold,kao2002subthreshold}. Consequently, \ieycr{2}{the} \maj{X} operation exhibit higher success rates \omcr{2}{at higher} temperature\omcr{2}{s}.

\observation{\ieycr{2}{Storing multiple copies of \maj{X}'s input operands} lowers the effect of temperature on success rate.} 
For example, performing \maj{3} with 32-row activation has at most \param{1.65}\% success rate \ieycr{0}{variations}, whereas \maj{3} with 4-row activation has at most \param{15.20}\%.

\begin{figure}[ht]
\centering
\includegraphics[width=\linewidth]{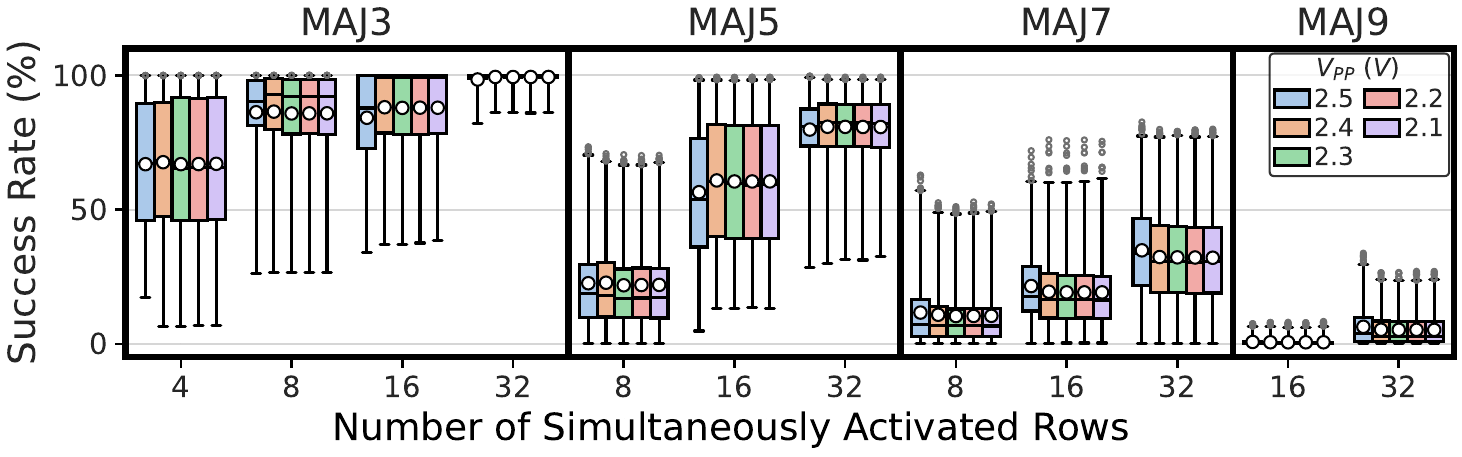}
\caption{\ieycr{2}{E}ffect of wordline voltage on the success rate of \ieycr{2}{the} \maj{X} operation.}
\label{fig:voltage_maj}
\end{figure}

\head{Voltage Scaling}
\figref{fig:voltage_maj} shows the success rate distribution of \ieycr{2}{the} \maj{X} operation \ieycr{0}{at} five \ieycr{0}{different} wordline voltage (\vpp{}) levels: 2.5V, 2.4V, 2.3V, 2.2V, and 2.1V.\footref{fn:voltage_fn} \ieycr{0}{We make Observation 13 from \figref{fig:voltage_maj}.}

\observation{\omcr{2}{Wordline v}oltage slightly affects the success rate of \ieycr{2}{the} \maj{X} operation.}
 \ieycr{0}{S}uccess rate \ieycr{0}{of the \maj{X} operation has a} \param{1.10}\% \ieycr{0}{variation} on average across all the tested operations. 

\ieycr{0}{We derive Takeaway 5 from Observations 9, 11, and 13.}
 \takeaway{\ieycr{0}{\omcr{2}{W}ordline voltage and temperature only slightly affect the success rate of the \maj{X} operation, whereas data pattern has a significant effect on success rate.}}

\setcounter{version}{1}
\section{\gf{Characterizing Multi-Row Copy}}
We experimentally characterize the \mrc{} operation where one \ieycr{0}{source} row's content can \omcr{0}{concurrently} be copied \mel{to} up to 31 \ieycr{0}{destination} rows.~We evaluate the robustness of the \mrc{} operation under four key parameters: \one{} timing delay between \ieycr{0}{each} command \ieycr{0}{pair} in \ieycr{0}{the} \apa{} \ieycr{0}{sequence}, \two{} data pattern, \three{} \ieycr{0}{chip} temperature, and \four{} wordline voltage. 
\label{sec:mrc_char}

\begin{figure}[ht]
\centering
\includegraphics[width=\linewidth]{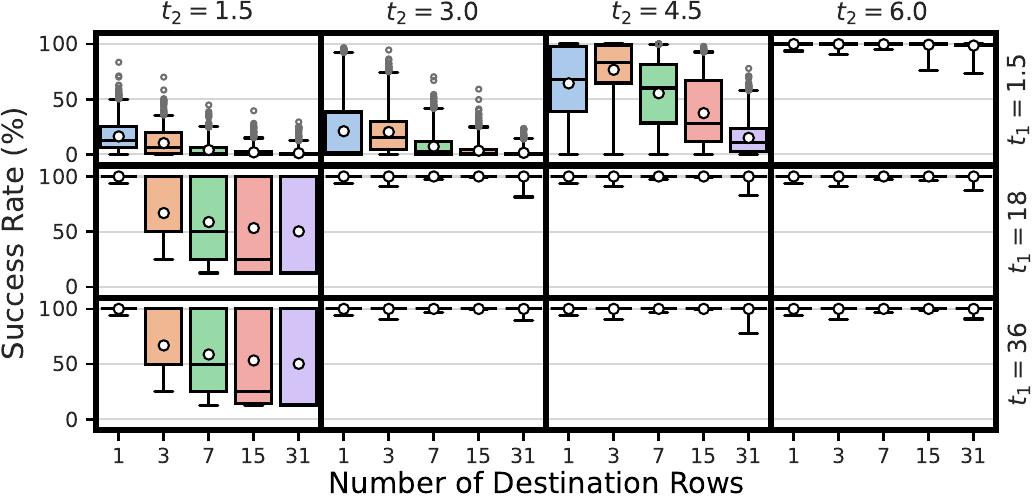}
\caption{Effect of $t_{1} (ns)$, $t_{2} (ns)$ and the number of simultaneously activated rows on the success rate of \ieycr{0}{the} \mrc{} operation.}
\label{fig:timing_mri}
\end{figure}

\head{Effect of Timing Delay}
\figref{fig:timing_mri} shows the distribution of the success rate for different numbers of \ieycr{1}{destination rows} in a box and whiskers plot\footref{fn:boxplot}
for different combinations of $t_{1}$, the timing between \act{} and \pre{} and $t_{2}$, the timing delay between \pre{} and \act{}. \ieycr{0}{We make Observations 14 and 15 from \figref{fig:timing_mri}.}

\observation{\ieycr{0}{COTS} DRAM chips can copy one row's content to up to 31 rows \ieycr{1}{with a} >\param{99.98}\% success rate.}
When we use the timing delays that achieve the highest success rate (i.e., $t_{1}$=36ns and $t_{2}$=3ns), copying one \ieycr{0}{source} row's content to 1, 3, 7, 15, and 31 \ieycr{0}{destination} rows has an average success rate of 
\param{99.99}\ieycr{1}{6}\%, \param{99.98}\ieycr{1}{9}\%, \param{99.99}\ieycr{1}{8}\%, \param{99.99}\ieycr{1}{9}\%, and \param{99.98}\ieycr{1}{2}\%.
We hypothesize that waiting for the \gls{tras} timing parameter (i.e., $t_{1}$=36ns) before issuing \ieycr{0}{a} \pre{} in \ieycr{0}{the} \apa{} command sequence ensures \omcr{1}{that} the sense amplifier senses the source row correctly and drives \omcr{1}{the} bitlines to the source row's charge \omcr{1}{level}.
\observation{Low $t_1$ (i.e., $t_1=1.5ns$) \mel{has} a significantly lower success rate than other timing configurations.}
When we choose $t_{1}$ as 1.5ns, \mrc{} achieves \param{49.79}\% lower success rate than the second worst timing configuration.
We hypothesize that decreasing $t_{1}$ (e.g., to $t_1=1.5ns$) could \omcr{1}{cause} the sense amplifiers to not fully drive the bitlines with the source row's charge\ieycr{0}{, which results in low success rate\omcr{1}{s}.}

\begin{figure}[ht]
\centering
\includegraphics[width=\linewidth]{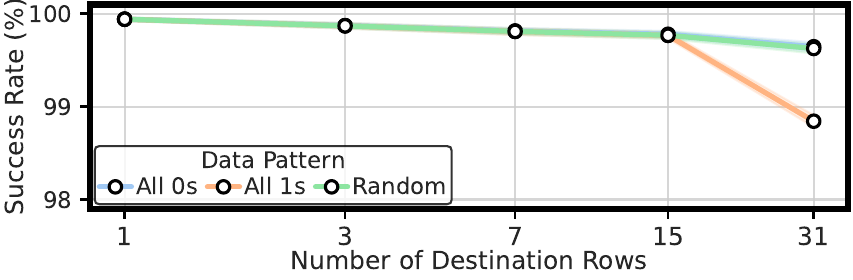}
\caption{Data pattern dependence of \mrc{}.}
\label{fig:data_mri}
\end{figure}

\takeaway{\ieycr{0}{COTS} DRAM chips are capable of copying one row's data to 1, 3, 7, 15, and 31 \ieycr{0}{other} rows \ieycr{0}{at very high success rates}.}
\head{Data Pattern}
\figref{fig:data_mri} shows the average success rate of \mrc{} (y-axis) for different numbers of \ieycr{1}{destination} rows (x-axis) with three data patterns (in different line colors). \ieycr{0}{We make Observation 16 from \figref{fig:data_mri}.}
\observation{Copying all-1s to 31 rows has a slightly lower success rate than copying other data patterns.}
Although copying up to 15 rows has a \ieycr{0}{small} (i.e., at most \param{0.11}\%) success rate difference among tested data patterns, when we copy all-1s to 31 rows, we observe a \param{0.79}\% decrease in success rate compared to all-0 and random data patterns. 

\begin{figure}[ht]
\centering
\begin{subfigure}[b]{0.5\linewidth}
\centering
\includegraphics[width=\linewidth]{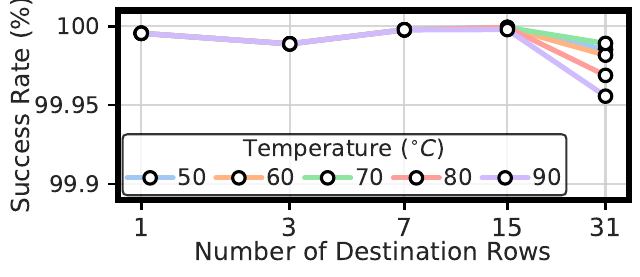}
\caption{\vspace{-1pt}}
\label{subfig:temp_mri}
\end{subfigure}
\begin{subfigure}[b]{0.5\linewidth}
\centering
\includegraphics[width=\linewidth]{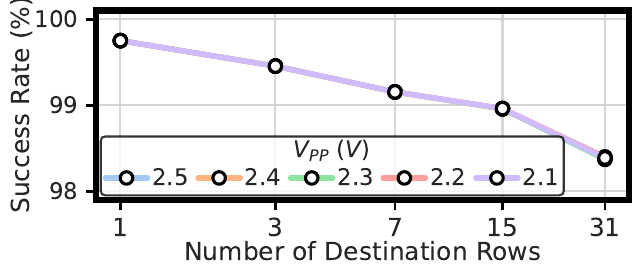}
\caption{}
\label{subfig:voltage_mri}
\end{subfigure}
\caption{
Average success rate of \mrc{} operations with \ieycr{0}{varying} (a) temperature and (b) wordline voltage scaling.}
\label{fig:temp_voltage_mri}
\end{figure}

\head{Temperature Scaling}
\figref{subfig:temp_mri} shows the average success rate of \mrc{} (y-axis) for different numbers of \ieycr{1}{destination} rows (x-axis) \ieycr{0}{at} five temperature levels (in different line colors): \SI{50}{\celsius}, \SI{60}{\celsius}, \SI{70}{\celsius}, \SI{80}{\celsius}, and \SI{90}{\celsius}. \ieycr{0}{We make Observation 17 from \figref{subfig:temp_mri}.}

\observation{Increasing temperature up to \SI{90}{\celsius} has a very small effect on the success rate of all the tested \mrc{} operations.}
Increasing the \ieycr{0}{DRAM chip} temperature from \SI{50}{\celsius} to \SI{90}{\celsius} has \param{0.04}\% success rate \ieycr{0}{variation} on average of all the tested \mrc{} operations.

\head{Voltage Scaling}
\figref{subfig:voltage_mri} shows the average success rate of \mrc{} (y-axis) for different numbers of \ieycr{1}{destination} rows (x-axis) \ieycr{0}{at} five voltage levels (in different line colors): 2.5V, 2.4V, 2.3V, 2.2V, and 2.1V.\footref{fn:voltage_fn} \ieycr{0}{We make Observation 18 from \figref{subfig:voltage_mri}.}

\observation{\ieycr{0}{Reducing} the wordline voltage has a \ieycr{0}{small} impact on the success rate.}
Underscaling the \vpp{} by 0.4V decreases the success rate by 1.32\% at most across all the tested \mrc{} operations.

\ieycr{0}{We derive Takeaway 7 from Observations 16-18.}
\takeaway{\mrc{} \ieycr{0}{in COTS DRAM chips} is highly resilient to changes in data pattern, temperature, and wordline voltage.}

\setcounter{version}{1}
\section{{\gf{\gls{pud} Operations in \ieycr{0}{COTS} DRAM \ieycr{0}{Chips}:\break Hypothesis \& Underlying Mechanisms}}}

\gf{We \omcr{1}{have} experimentally demonstrate\omcr{1}{d} that \ieycr{0}{COTS} DRAM chips \ieycr{0}{are capable of performing} \gls{pud} operations by violating DRAM timing parameters and leveraging simultaneous many-row activation. 
To fundamentally understand \omcr{1}{the reasons of this capability}, one would have to obtain access to the micro architectural design of the DRAM modules we test, which, unfortunately, is \emph{not} publicly available. 
However, based on the known literature on modern DRAM chips\ieycr{0}{~\cite{gao2019computedram,gao2022frac,olgun2021quactrng,dram-circuit-design,weste2015cmos,bai2022low,turi2008high,olgun2023dram}}, open-sourced SPICE models of DRAM arrays\ieycr{0}{~\cite{hassan2016chargecache,chang2017understanding,luo2020clrdram}}, and our observations \ieycr{0}{(\secref{sec:sim-act}-\ref{sec:mrc_char})}, we draw two hypotheses to shine light on our experimental observations.~First, we hypothesize that simultaneous many-row activation is possible in real DRAM chips due to the \emph{hierarchical DRAM row decoder} modern DRAM devices employ \ieycr{0}{(\secref{sec:reverse_engineering})}.
Second, we hypothesize that input replication improves the success rate of \maj{X} operation by increasing bitline voltage perturbation toward a margin that the DRAM sense amplifier can produce correct \maj{X} result \ieycr{0}{(\secref{sec:maj_hypo})}.}

\subsection{Hypothetical Row Decoder Design}
\label{sec:reverse_engineering}
\atb{We explain how \gf{a} hierarchical DRAM row decoder circuitry~\cite{weste2015cmos}\gf{,} \omcr{1}{which is} used to construct high-performance and high-density DRAM chips\gf{,} could allow for simultaneous activation of many \gf{DRAM} rows.}
The row decoder circuitry in a DRAM bank decodes the n-bit row address (RA) and asserts a wordline out of $2^n$ wordlines. Modern DRAM chips have multiple tiers of decoding stages to reduce latency, area, and power consumption~\cite{bai2022low,weste2015cmos,turi2008high}. We analyze the row decoder circuitry of a \omcr{1}{COTS} DRAM \hluo{chip}~\cite{hynix-reveng-module}, which has \param{$2^{16}$} rows in a bank. We observe that in this chip, each subarray consists of \param{$2^9$} rows and the number of subarrays in a bank is \param{$2^7$}. We present a hypothesis regarding the row decoder circuitry \yct{that allows simultaneous many-row activation} and the sequence of operations that occur in the row decoder when \omcr{1}{consecutive} \act{} and \pre{} commands are issued.

\begin{figure}[ht]
\centering
\includegraphics[width=\linewidth]{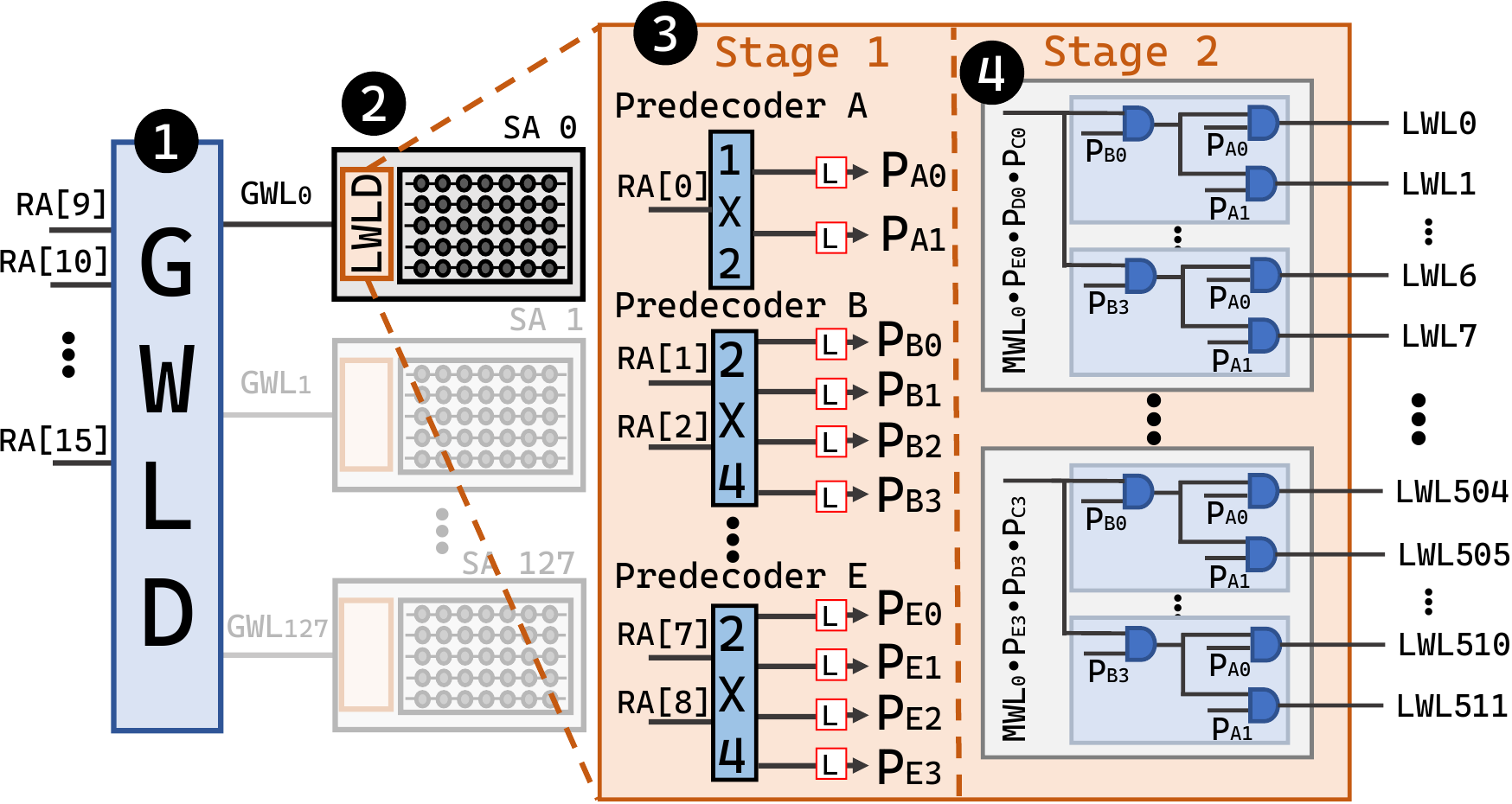}
\caption{Hypothetical row decoder design.}
\label{fig:row-decoder}
\end{figure}

\noindent
\textbf{Row Address Indexing.} \atb{We observe that} the lower-order 9 bits of the RA are used to index \atb{a row in} a subarray, while the higher-order 7 bits are used to index \atb{a} subarray \atb{in a bank}. \atb{To do so, we carefully reuse the DRAM row adjacency reverse engineering methodology described in prior works~\cite{orosa2021deeper,yaglikci2022hira}}. 
\noindent
\textbf{\atb{Detailed} Row Decoder Design.}~\figref{fig:row-decoder} illustrates \atb{a hypothetical} row decoder circuitry \atb{in} a bank
that consists of two \atb{components}: 1) Global Wordline Decoder (GWLD) (\dingOne{}) and 2) Local Wordline Decoder (LWLD) (\dingTwo{}). When an \act{} command is issued, three operations occur in \atb{the following} order. First, GWLD decodes the higher-order 7 bits of the RA (RA[9:15]) and drives the Global Wordline ($GWL$) that \atb{enables} the LWLD of \yct{\atb{a}} subarray \atb{(e.g., GWL0 enables LWLD of SA0 in~\figref{fig:row-decoder})}. Second, \yct{Stage 1 of LWLD pre}decodes the lower-order 9 bits of the RA \yct{(RA[0:8])} in five tiers of predecoders \yct{(Predecoder A, B, C, D, and E; \dingThree{})} and latches the predecoded address bits \yct{($P_{A0}, P_{A1}, ..., P_{E3}$)}. Third, \yct{Stage 2 of LWLD decodes the} predecoded $P$ signals to assert the corresponding Local Wordline ($LWL$)~\yct{in} Stage 2 (\dingFour{}). When a \pre{} command is issued \atb{with standard DRAM timing parameters}, the latched predecoded \yct{$P$ signals} are \atb{correctly} \yct{de-asserted} the corresponding $LWL$.

\noindent
\textbf{Activating Multiple Rows: A Walk-Through.}
Reducing the latency between \pre{} and the \yct{second} \act{} commands (i.e., $t_{RP}$) allows the predecoders to latch the next RA without deasserting the RA \yct{targeted by} the first \act{} command. Hence, after the second \act{} command, depending on the target addresses of \apa{} sequence, \ieycr{1}{multiple} latches of \yct{each} predecoder in LWLD \yct{can be} set.
\yct{By changing the row addresses targeted by two \act{} commands, we can control the number and addresses of the simultaneously activated rows in a subarray.} 

\begin{figure}[ht]
\centering
\includegraphics[width=\linewidth]{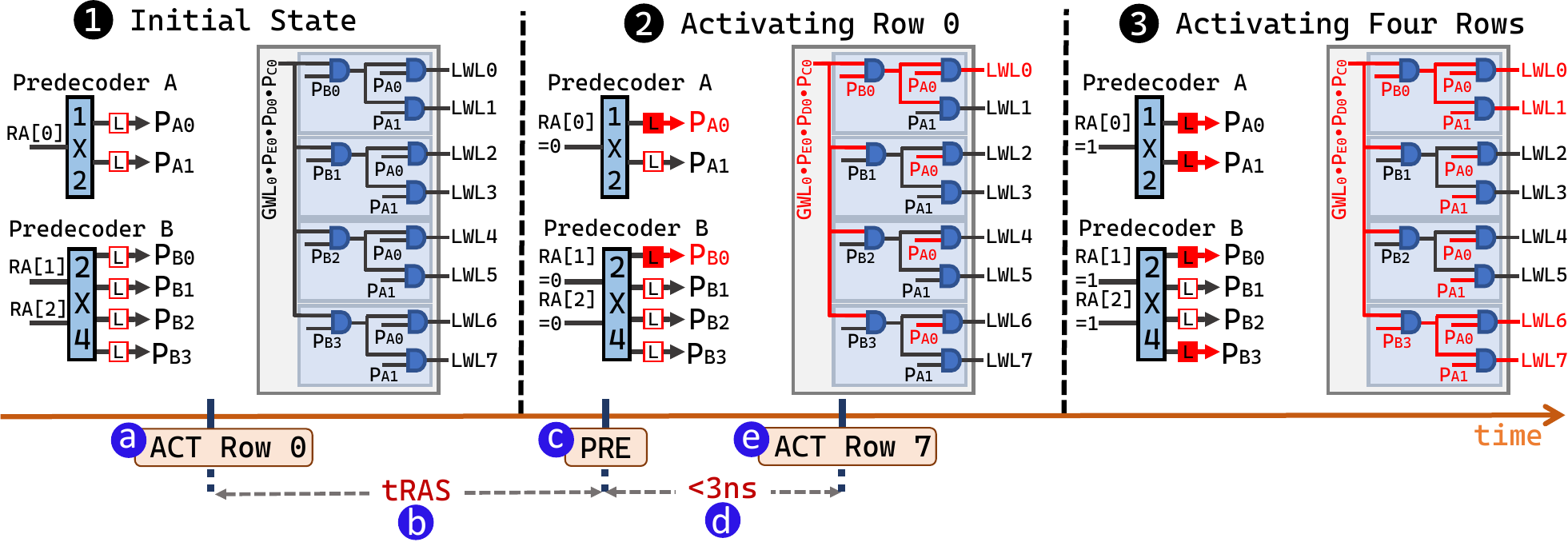}
\caption{Example of activating multiple rows in hypothetical row decoder design. \ieycr{1}{R}ed represent\ieycr{1}{s} asserted signals.}
\label{fig:mra_example}
\end{figure}

\yct{\figref{fig:mra_example} demonstrates an example of simultaneously activat\atb{ing} four rows in the same subarray
when the \apaex{} is issued \ieycr{1}{with violated timings}.}\ieycr{1}{The memory controller issues each command (shown in orange boxes below time axis) at the corresponding tick mark, and asserted signals are highlighted in red. The bank is initially precharged and no signals in the row decoder circuitry are asserted (\dingOne{}).} 

\ieycr{1}{We simultaneously activate four rows in X steps. First, we issue an \act{} command to \texttt{Row 0} (\dingA{}), and wait for the manufacturer-recommended \gls{tras} (\dingB{}). This results in the predecoders asserting $P_{A0}$ and $P_{B0}$ signals, which drive $LWL_{0}$ and activate \texttt{Row 0} (\dingTwo{}). Second, we issue a \pre{} command (\dingC{}) with violated \gls{trp} timing, e.g., $<$3ns (\dingD{}), and we issue another \act{} command to \texttt{Row 7} (\dingE{}). Issuing back-to-back \pre{}$\rightarrow$\act{} asserts $P_{A1}$ and $P_{B3}$ signals without de-asserting $P_{A0}$ and $P_{B0}$ (\dingThree{}).} As a result, $P_{A0}$, $P_{A1}$, $P_{B0}$, and $P_{B3}$ signals are set simultaneously, \agy{and thus the} decoder tree asserts $LWL_{0}$, $LWL_{1}$, $LWL_{6}$, and $LWL_{7}$ \agy{wordlines}, \agy{thereby} simultaneous\agy{ly} activating rows 0, 1, 6, and 7.

We formulate our observation as follows: to activate $2^N$ rows, N different predecoders have to latch two predecoded P signals. For instance, as illustrated in \figref{fig:mra_example}, to activate 4 rows, we issue \apa{} command \ieycr{2}{sequence that targets} the rows that only latch the \yct{two different predecoders}' (i.e., Predecoders A and B) two outputs (i.e., \{$P_{A0}$, $P_{A1}$\} and \{$P_{B0}$, $P_{B3}$\}). Hence, to activate 32 rows, an \apa{} sequence needs to target such rows that make all predecoders latch two outputs (e.g., \act{} 127 $\rightarrow$ \pre{} $\rightarrow$ \act{} 128). We hypothesize that the upper bound for the number of rows that are simultaneously activated depends on the number of predecoders. The examined module \atb{likely} has five predecoders, and thus, we can activate up to $2^5$ rows.

\subsection{Increasing Success Rate by Leveraging Input Replication}

\label{sec:maj_hypo}
Current DRAM standards do not officially support the \maj{X} operation. Yet, by leveraging the fundamental design and operation\omcr{1}{al principles} of COTS DRAM chips, it is possible to perform \maj{X} by issuing \apa{} with violated timing parameters: \gls{tras} and \gls{trp}~(\secref{sec:maj_char}).
\omcr{1}{Doing so can reduce} \ieycr{1}{the bitline voltage perturbation compared to a standard DRAM activation operation (i.e., single row activation)}, as multiple cells are \ieycr{1}{simultaneously} connected \ieycr{1}{to the same bitline}. \ieycr{1}{As a result, the reduced bitline voltage perturbation is less likely to exceed reliable sensing margin, and the sense amplifier fails to reliably produce correct \maj{X} results.} We hypothesize that by storing multiple copies of \maj{X} input \ieycr{1}{operands} on all simultaneously activated rows (which we call input replication), the bitline voltage can be increased and perturbed towards a safer margin and, thus, potentially increase the success rate of \maj{X} operations.

\begin{figure}[ht]
\centering
\begin{subfigure}[b]{0.45\linewidth}
\centering
\includegraphics[width=\linewidth]{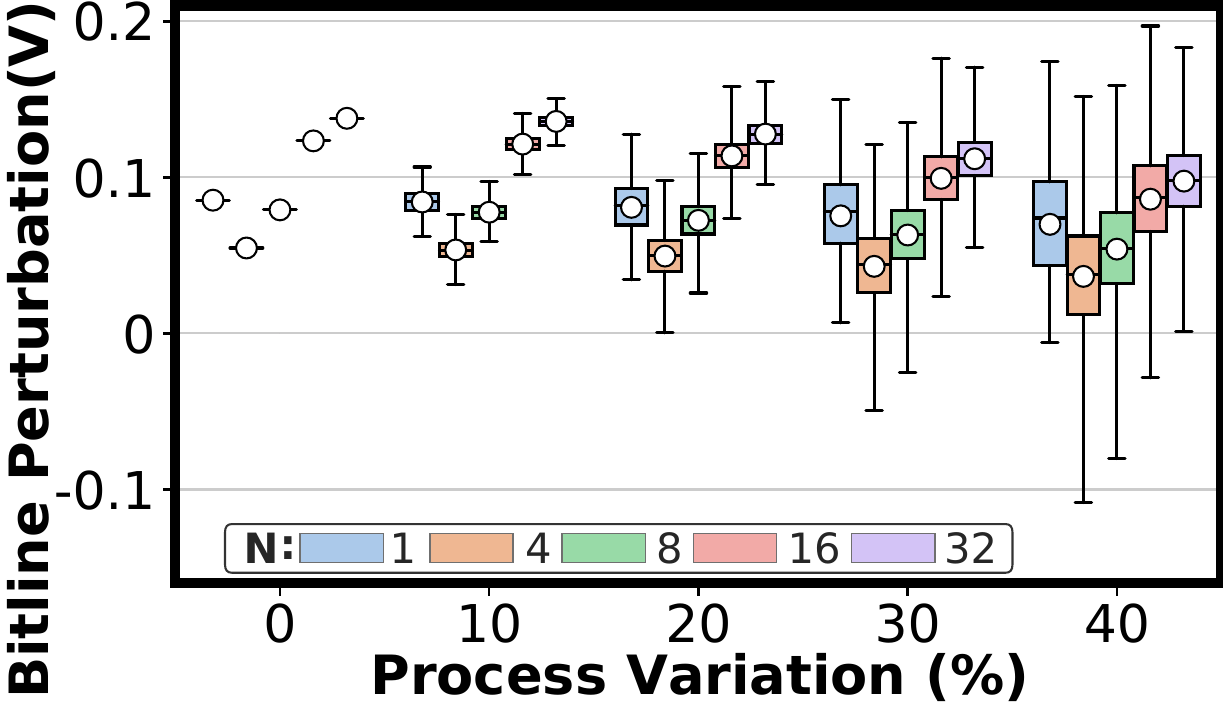}
\caption{}
\label{subfig:repl_spice2}
\end{subfigure}
\begin{subfigure}[b]{0.45\linewidth}
\centering
\includegraphics[width=\linewidth]{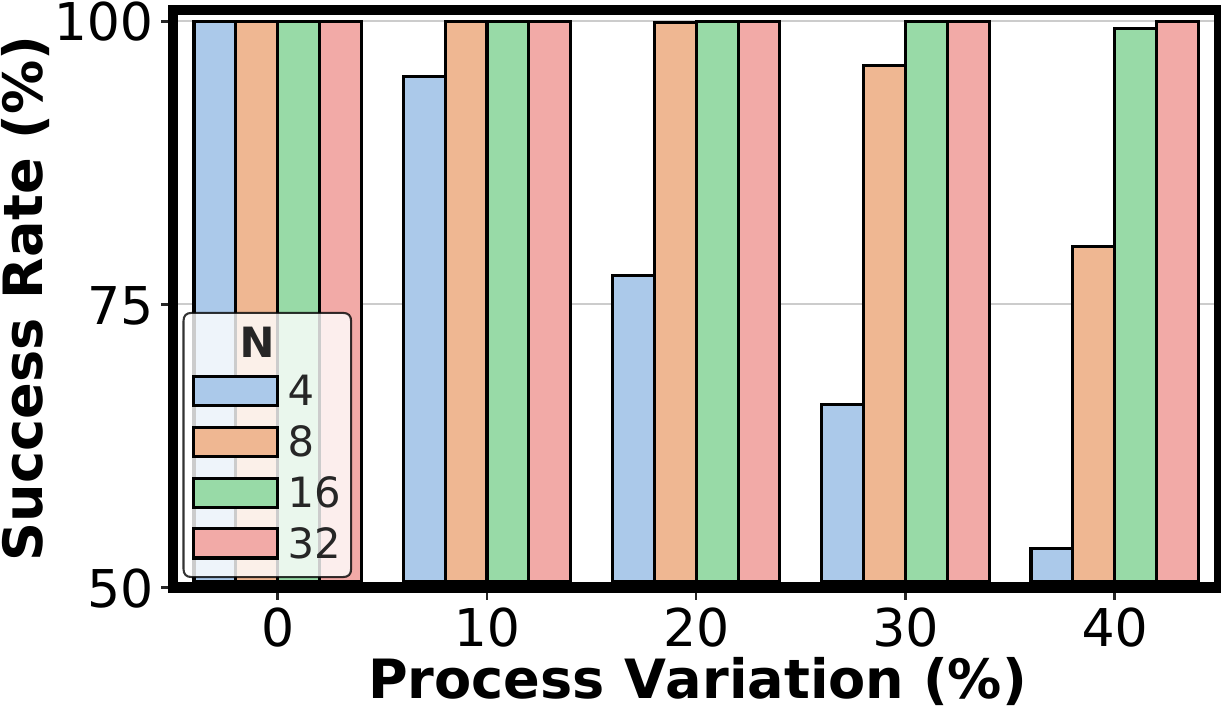}
\caption{}
\label{fig:spice_all}
\end{subfigure}
\caption{(a) Effect of input replication on the bitline deviation (b) and the success rate of \maj{3} for N-row activation across different process variations using SPICE simulations}
\label{fig:repl_spice}
\end{figure}

To \omcr{1}{test} our hypothesis, we conduct SPICE simulations and analyze the effect of input replication on the success rate of \ieycr{1}{one example \maj{X} operation, \maj{3}(1,1,0).} 
\figref{fig:repl_spice} shows the effect of process variation on the sensing operation for \maj{3}(1,1,0) with $N$-row activation, where $N \in\{1,4,8,16,32\}$.
\figref{subfig:repl_spice2} depicts the \ieycr{1}{bitline perturbation} distribution right before the sensing operation (y-axis) across 1000 different sets of $N$ DRAM cell(s) (e.g., for 4-row activation, we test 1000 different sets of four cells) for different process variation percentages (x-axis).
Each $N=1$ box represents the bitline \ieycr{1}{bitline perturbation} distribution for a single-row activation. Boxes for other $N$ values show the \ieycr{1}{bitline perturbation} distribution for 4-, 8-, 16-, and 32-row activation.
\figref{fig:repl_spice}b shows the success rate of \maj{3}(1,1,0) with $N$-row activation, where $N \in\{4,8,16,32\}$.

We make three key observations based on \figref{fig:repl_spice}. First, increasing the number of simultaneously activated rows increases the \ieycr{1}{bitline perturbation} in every process variation percentage. On average, performing \maj{3} with 32-row activation (i.e., ten copies for each input operand) has \param{159.05}\% higher \ieycr{1}{bitline} voltage \ieycr{1}{perturbation} than performing \maj{3} with 4-row activation on average. Second, activating more than eight rows always results in a higher \ieycr{1}{bitline perturbation} than single-row activation on average for every process variation percentage. 
\nb{Third, input replication results in a higher success rate under \yct{all} process variation \yct{percentages}. The success rate of} \maj{3} \nb{with 4-row activation reduces by \param{46.58}\% when process variation increases from 0\% to 40\%. In contrast, the success rate of }\maj{3} \nb{with 32-row activation reduces only by \param{0.01}\%.} 
We conclude that \omcr{1}{higher} success rate\omcr{1}{s observed with replicated inputs in COTS DRAM chips (\secref{sec:maj_char}) due to bitline voltage is perturbed towards a more reliable sensing margin than no replication.}

\setcounter{version}{1}

\section{Case Studies}
We study the potential \omcr{1}{performance} benefits of enabling \omcr{1}{the} new \gls{pud} operations (i.e., \maj{5}, \maj{7}, \maj{9}, and \mrc{}) \omcr{1}{we demonstrated} in COTS DRAM chips. We analyze the potential performance gain using new \gls{pud} operations in \one{} seven majority-based arithmetic \& logic operations and \two{} content destruction operations for cold-boot attack prevention.
\subsection{Majority-Based Computation}

\head{Majority-Based Boolean and Arithmetic Operations}
Majority operation\mel{s} can be used to implement 1) logic operations such as AND/OR~\cite{hajinazar2020simdram,ali2019memory,seshadri2016buddy,seshadri2017ambit,seshadri2015fast,seshadri2019dram,gao2019computedram}) and XOR~\cite{alkaldy2014novel}, and 2) full add\omcr{1}{ition}~\cite{ali2019memory,gao2019computedram,hajinazar2020simdram,deoliveira2024mimdram}. These operations \omcr{1}{can be} used as basic building blocks for \omcr{1}{more complex} \gls{pud} computation (e.g., multiplication)~\cite{ali2019memory,angizi2019graphide,li2016pinatubo,gao2019computedram}.

\head{Experimental Methodology}
\nb{We evaluate the execution time of seven arithmetic \& logic microbenchmarks implemented using \maj{X} operations (\maj{3}, \maj{5}, and \maj{7} for Mfr. M and \maj{3}, \maj{5}, \maj{7}, and \maj{9} for Mfr. H). We perform 32-bit logic (\mel{AND, OR, and XOR}) and arithmetic (addition, subtraction, multiplication, and division) computations on 8KB elements.}

\nb{We use DRAM Bender~\cite{olgun2023dram} to tightly schedule the DRAM commands to perform \maj{X}, \mrc{}, and RowClone operations and measure their latency. To measure the latency of \maj{X} operations with N-row activation, we perform RowClone to copy the \maj{X} inputs into X number of rows and replicate the input operations into N rows using \mrc{} operations. We use the empirical success rates (obtained from DRAM chips) and calculate the throughput for each \maj{X} operation. We then choose the group of rows that are simultaneously activated, which produces the highest throughput across all \nMODULES{} tested DRAM modules for each \maj{X} operation. We analytically model the execution time of the arithmetic and logic microbenchmarks using the highest throughput values.}
\nb{For the baseline, we use the highest throughput of \maj{3} with 4-row activation and RowClone operations, which is the state-of-the-art for \gls{pud} operations.}

\begin{figure}[ht]
\centering
\includegraphics[width=0.85\linewidth]{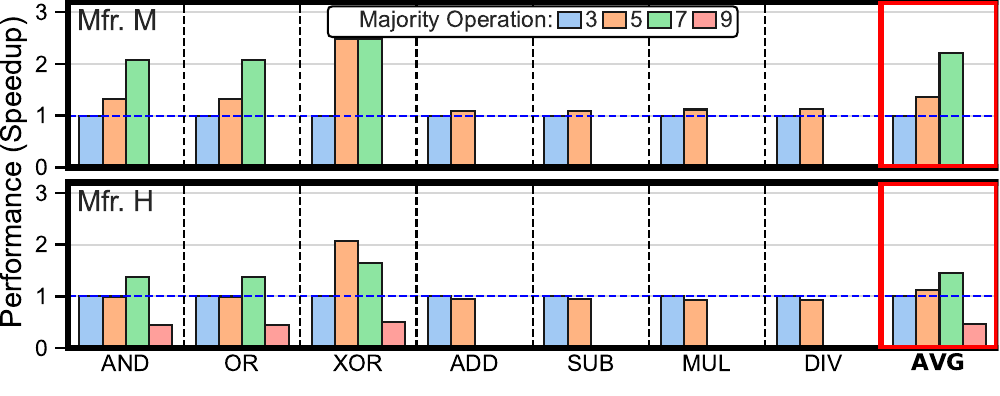}
\caption{Speedup \omcr{1}{of using \maj{5}, \maj{7}, and \maj{9}} over the state-of-the-art (\maj{3}) in \yct{seven} arithmetic \& logic microbenchmarks.}
\label{fig:maj_perf_1}
\end{figure}

\head{\nb{Performance Evaluation}} \figref{fig:maj_perf_1} shows the execution time \omcr{1}{speedup} of \nb{seven microbenchmarks that use \maj{X} operations in DRAM chips from two manufacturers} normalized to the baseline (i.e., \maj{3} with 4-row activation), the blue dashed line. We make three key observations. First, new \maj{X} operations (i.e., \maj{5}, \maj{7}, and \maj{9}) \omcr{1}{improve performance over} \maj{3} in all microbenchmarks. On average, new \maj{X} operations provide \param{121.61\%} (\param{46.54}\%) higher performance over using \nb{only} \maj{3} in Mfr. M (Mfr. H). Second, increasing the number of input operands in \maj{X} \nb{improves} performance. \maj{7} provides \param{62.10\%} (\param{31.71\%}) \omcr{1}{higher performance than} \maj{5} in Mfr. M (Mfr. H). Third, in Mfr. H, \maj{9} \omcr{1}{leads to performance degradation of 114.12\%.} This is because \maj{9} has a poor success rate (shown in \figref{fig:data_maj}), which requires repeatedly performing the \maj{9}, resulting in higher \nb{execution time}. We conclude that enabling new \maj{X} operations has great potential to improve the performance of majority-based computation.

\head{Majority-based Error Correction Operations}
Majority operations could be useful for many systems that experience errors frequently, such as systems in space environments~\cite{wirthlin2016seu,jacobs2012reconfigurable,siegle2015mitigation}. One common reliability technique for mitigating these errors is triple modular redundancy (TMR)~\cite{wirthlin2016seu,lyons1962use}. TMR stores three copies of the original data and performs majority voting to decide the correct output. A single \omcr{1}{bit} error in TMR \omcr{1}{does} not produce any error as the majority operation \omcr{1}{would calculate} the correct result \omcr{1}{based on the other two correct copies of data}~\cite{wirthlin2016seu,lyons1962use}.

\maj{3} operations can be leveraged to perform majority voting and thus could be useful for such systems. Since we observe that off-the-shelf DRAM chips can perform up to seven input majority operations (i.e., \maj{9}), \maj{X} operations can be used to correct not only a single fault but up to three faults in the systems. Due to space limitations, we leave the exploration of this use case to future work.

\subsection{Content Destruction-Based Cold-Boot Attack\\ Prevention}
\head{Cold-Boot Attack Prevention Mechanism} 
A cold boot attack is a physical attack on DRAM that involves hot-swapping a DRAM chip and reading out the contents of the DRAM chip ~\cite{bauer2016lest,gruhn2013practicality,halderman2009lest,hilgers2014post,lee2012correcting,lindenlauf2015cold,muller2010aesse,simmons2011security,villanueva2019cold,yitbarek2017cold}. Cold boot attacks are possible because the data stored in DRAM is not immediately lost when the chip is powered off. This is due to the capacitive nature of DRAM cells that can hold their data up to several minutes~\cite{halderman2009lest,bauer2016lest,khan2014efficacy,liu2013experimental,liu2012raidr,patel2017reaper}. A practical and secure way to mitigate cold boot attacks is to destroy the DRAM content rapidly during power-off/on~\cite{orosa2021codic,tcg2008platform}. Off-the-shelf \gls{pud} operations (i.e., RowClone~\cite{gao2019computedram,seshadri2013rowclone, olgun2021pidram}, \texttt{Frac}~\cite{gao2022frac}, and \mrc{} \omcr{1}{(\secref{sec:mrc_char})}) can be used to rapidly destroy the DRAM content. 

\head{Experimental Methodology} 
We schedule \nb{DRAM command sequences} to perform all content destruction operations (i.e., RowClone, \texttt{Frac}, and \mrc{}) and measure the latency of each operation using DRAM Bender~\cite{olgun2023dram}. We then use an analytical model to calculate the total time to overwrite all data in a DRAM bank. The three \gls{pud} operation-based content destruction operations (i.e., RowClone-based, \texttt{Frac}-based, and \mrc{}-based) can be implemented as follows: \one{} \textbf{RowClone-based content destruction} \omcr{1}{is} a two-step process. First, we issue a \wri{} command to write predetermined data to an arbitrary row. Second, we perform RowClone \omcr{1}{from that row to all other DRAM rows, rewriting all} original content. \two{} \textbf{\texttt{Frac}-based content destruction} is implemented to repeatedly send the \texttt{Frac} operation to every row to \omcr{1}{place all DRAM} rows into a neutral state, making them store \vddh{}. \three{} \textbf{\mrc{}-based content destruction} \omcr{1}{is} implemented with varying numbers of rows that are simultaneously activated, from 2 to 32, in two steps. First, we issue a \wri{} command to write predetermined data to an arbitrary row. Second, we perform up to 32-row activation with \mrc{} operation to rapidly destroy the data in all rows.

\begin{figure}[ht]
\centering
\includegraphics[width=0.78\linewidth]{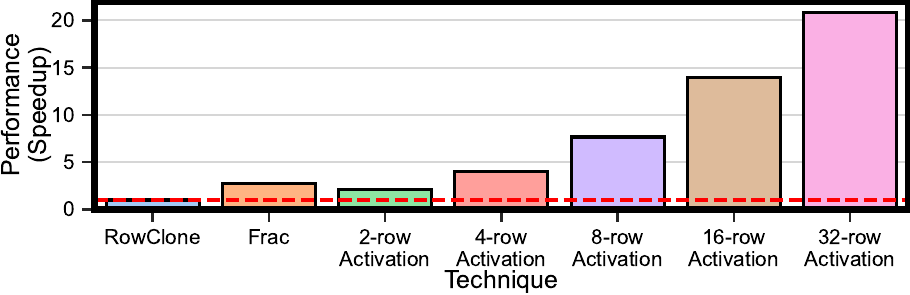}
\caption{Speedup over RowClone-based content destruction}
\label{fig:cold_boot}
\end{figure}

\head{Performance Evaluation} \figref{fig:cold_boot} \agy{shows} the speedup in execution time for content destruction normalized to the RowClone-based content destruction's execution time. 

\nb{We make two key observations \agy{based on \figref{fig:cold_boot}}. 
First, \mrc{}-based content destruction with 4-, 8-, 16- and 32-row activation outperforms both RowClone-based and \agy{\texttt{Frac}}-based content destruction \omcr{1}{by} up to \param{20.87}$\times$ and \param{7.55}$\times$, respectively. 
Second, increasing the number of simultaneously activated rows increases the speedup of \omcr{1}{the} \mrc{}\-/based technique\mel{, as }increasing the number of operands in \mrc{} decreases the total number of operations.}
We conclude that \mrc{}\nb{-based content destruction has a great potential to \omcr{1}{enable rapid destruction of} \agy{DRAM} content.}

\setcounter{version}{1}
\section{Limitations \ieycr{1}{of Tested COTS DRAM Chips}}
\label{sec:limitations}
\ieycr{0}{We identify three key limitations of COTS DRAM chips in performing \gls{pud} operations.\footnote{We provide more details and limitations that we identify in the extended version~\cite{yuksel2024simultaneous}.}}

\head{\ieycr{0}{Limitation 1. Some COTS DRAM chips do \emph{not} support all \gls{pud} operations}}
While we test COTS DRAM chips from all three major manufacturers (i.e., SK Hynix, Samsung, and Micron), we report major positive results and detailed evaluations in DRAM chips from Micron (Mfr. M) and SK Hynix (Mfr. H), which share more than half of the DRAM market (i.e., 55.8\%)~\cite{trendforce2023market}. 
\ieycr{0}{we can \emph{simultaneously} activate many rows in a subarray, and thus, we can perform \emph{all} the tested \gls{pud} operations. In the tested Samsung chips\footnote{We provide more details on every tested chip (including Samsung chips) in the extended version of this paper~\cite{yuksel2024simultaneous}.} (a total of 64 DRAM chips), we do not observe simultaneous activation of more than one row in a subarray. Hence, we do not observe any tested \gls{pud} operations in Samsung chips.} We hypothesize that these DRAM chips have internal circuitry that ignores the \pre{} command or the second \act{} command when the timing parameters (\gls{trp} and \gls{tras}) are greatly violated, which \nb{is in line} with the hypotheses provided by prior work~\cite{yaglikci2022hira}. \ieycr{0}{If such \omcr{1}{a} limitation \omcr{1}{were} not imposed, we believe these DRAM chips are also fundamentally capable of performing the operations we examine in this work.}

\head{\ieycr{0}{Limitation 2. Tested COTS DRAM chips support only consecutive activation of two rows and simultaneous activation of 2, 4, 8, 16, and 32 rows}}
\ieycr{0}{In our experiments, we observe that we cannot control how many rows that DRAM chips simultaneously activate: either we can consecutively activate two rows or simultaneously activate 2, 4, 8, 16, and 32 rows. We hypothesize that this is due to our current infrastructure limitation\omcr{1}{s}, where we can issue DRAM commands at intervals of only 1.5ns. Controlling \omcr{1}{the number of activated rows may} require finer-grain\omcr{1}{ed} timing control between DRAM commands (e.g., 0.1ns). Having fine-grain\omcr{1}{ed} control would allow us to deassert/assert desired intermediate signals in the row decoder circuitry, which could result in different numbers of simultaneously activated rows than 2, 4, 8, 16, and 32. }

\head{\ieycr{0}{Limitation 3. Performing \gls{pud} operations potentially have an effect on transient errors in DRAM chips}}
We perform each test 10000 times for each simultaneously activated row group and check for bitflips in the whole DRAM bank. We do not observe any errors in rows outside of the \omcr{1}{simultaneously activated} row group across any of the tested DRAM chips. We believe that investigating all potential effects \omcr{1}{(e.g., on transient errors} requires a much more extensive exploration of various aspects, which warrants its own study.

Our goal is to understand the undocumented computational capability of DRAM chips and the reliability of such computation capability. Despite the limitations mentioned above, our results can inspire future works to make PUD operations more reliable and/or leverage these operations in new ways in different applications with different requirements. We hope that future work builds on our study and that the resulting body of work enables DRAM manufacturers to \ieycr{1}{reliably} support operations that we demonstrate in future DRAM standards.

\glsresetall
\setcounter{version}{1}
\section{Related Work}
\label{sec:related-work}

To our knowledge, this is the first work that \ieycr{1}{\one{}} rigorously \ieycr{0}{characterizes the robustness of simultaneously activating up to 32 rows \ieycr{1}{\two{} demonstrates \maj{X} operations where X$>$3 \three{} demonstrated concurrently copying one row's content to up to 31 other rows (i.e., \mrc{}), and \four{} shows the success rate of \maj{X} can be improved by replicating the input \ieycr{0}{operand}s of \maj{X} in COTS DDR4 DRAM chips.} \ieycr{0}{We discuss related works in \ieycr{1}{two} synergistic directions: 1) \gls{pud} in COTS DRAM chips and 2) \gls{pud} in modified DRAM chips.}}

\subsection{\gls{pud} in COTS DRAM Chips}

\noindent\textbf{Multiple Row Activation-based \gls{pud} Operations.}
\omcr{1}{Several} prior works demonstrate bulk bitwise and data copy operations in COTS DRAM devices using multiple row activation~\cite{olgun2021quactrng,gao2019computedram,gao2022frac,yaglikci2022hira,olgun2023dram,yuksel2024functionallycomplete,olgun2021pidram}. 
\ieycr{0}{ComputeDRAM~\cite{gao2019computedram} \one{} activates three rows simultaneously (i.e., triple-row activation~\cite{seshadri2017ambit,seshadri.bookchapter17,seshadri2015fast,seshadri2016buddy,seshadri2016processing,seshadri2019dram}) {to perform} three-input majority and two-input AND and OR operations~\cite{seshadri2017ambit,seshadri.bookchapter17,seshadri2015fast,seshadri2016buddy,seshadri2016processing,seshadri2019dram} and \two{} demonstrates copying one row's content to another row (i.e., the RowClone~\cite{seshadri2013rowclone} operation) in COTS DDR3 chips. FracDRAM~\cite{gao2022frac} {shows that a DRAM cell in COTS DDR3 chips can store fractional values (e.g., VDD/2). FracDRAM uses fractional values to perform MAJ3 operations while simultaneously activating four DRAM rows in the same subarray.} 
DRAM Bender~\cite{olgun2023dram,safari-drambender} demonstrates two{-}input AND and OR operations in {COTS} DDR4 chips. PiDRAM~\cite{olgun2021pidram,olgun2021pidramgithub} provides a flexible end-to-end FPGA-based framework that enables {real system studies} of PuD techniques, such as RowClone~\cite{seshadri2013rowclone,gao2019computedram}. A concurrent work~\cite{yuksel2024functionallycomplete} demonstrates that COTS DRAM chips are capable of \one{} performing NOT and up to 16-input AND, NAND, OR, and NOR operations and \two{} simultaneously activating up to 48 DRAM rows in two neighboring subarrays. None of these works \one{} characterize the \omcr{1}{\emph{robustness}} of simultaneous many-row activation, \two{} demonstrate \maj{X}, where X$>$3, \omcr{1}{\three{} demonstrate} copying one row's content to up to 31 other rows (\mrc{}), and \four{} provide a detailed hypothetical row decoder design that explains how \omcr{1}{and why} many rows can be simultaneously activated.}

Other works enable different functionalities using simultaneous many-row activation. QUAC-TRNG~\cite{olgun2021quactrng} introduces quadruple row activation and exploits this phenomenon to generate true random numbers in off-the-shelf DRAM chips. \ieycr{0}{Our observations (e.g., \mrc{} and simultaneously activating up to 32 DRAM rows) could also be leveraged to generate true random numbers. HiRA~\cite{yaglikci2022hira} demonstrates {that real DRAM chips are capable of activating} two rows (in quick succession) in electrically isolated (i.e., \emph{not} physically adjacent) subarrays (called \emph{hidden row activation}). This work uses hidden row activation to parallelize a DRAM row's refresh operation with the refresh or activation of other rows in the same bank.}

\noindent
\textbf{Security Primitives.}
Prior works propose DRAM-based mechanisms to implement true random number generators (TRNGs) and physical unclonable functions (PUFs). DRAM-based TRNGs generate true random numbers by violating timing parameters~\cite{kim2019drange,talukder2019exploiting}, using \omcr{1}{data} retention failures~\cite{keller2014dynamic,sutar2016d} and using startup values~\cite{eckert2017drng,tehranipoor2016robust}. DRAM-based PUFs generate signatures by using retention-based failures~\cite{keller2014dynamic,sutar2016d,xiong2016run}, violating timing parameters~\cite{kim2018dram}, exploiting write access latencies~\cite{hashemian2015robust}, and using startup values~\cite{tehranipoor2016dram}.

\subsection{\ieycr{0}{PuD in Modified DRAM Chips}} 
\ieycr{0}{Prior works~\cite{li2016pinatubo,li2017drisa,Li2018SCOPEAS,manning2018apparatuses,oliveira2022accelerating,parveen2017hybrid,parveen2017low,parveen2018hielm,parveen2018imcs2,rakin2018pim,ramanathan2020look,rezaei2020nom,seshadri2016buddy,seshadri2013rowclone,seshadri2015fast,seshadri2015gather,seshadri2016processing,seshadri2017ambit,seshadri2017simple,seshadri2018rowclone,seshadri2019dram,shafiee2016isaac,song2017pipelayer,song2018graphr,tian2017approxlut,wu2022dram,xie2015fast,xin2019roc,xin2020elp2im,yang2020flexible,yu2018memristive,zawodny2018apparatuses,zha2020hyper,zhao2017apparatuses} enable bulk operations in DRAM chips by \emph{modifying} the DRAM circuitry. We demonstrate that COTS DRAM chips are capable of performing \gls{pud} operations \emph{without} any modification \omcr{1}{to DRAM chips or interface}. We believe truly \omcr{1}{and robustly} supporting operations that we demonstrate in DRAM requires {proper} modifications to DRAM circuitry and standards. Yet, our demonstration shows that existing COTS DRAM chips are already quite capable of computation, and such \omcr{1}{proper} modifications to DRAM are {very promising and} likely to be {very} fruitful.}

\setcounter{version}{1}
\section{Conclusion}
\ieycr{1}{We presented our extensive characterization study on the computational capability of COTS DRAM chips and the robustness of these capabilities in \nCHIPS{} COTS DDR4 DRAM chips. Our study leads to 18 new empirical observations and shares 7 key takeaway lessons, which demonstrate that COTS DRAM chips are capable of performing \maj{5}, \maj{7}, \maj{9}, and copying one row's content to up to 31 DRAM rows.} We believe that {our \omcr{1}{rigorous} empirical} results demonstrate the potential of using DRAM as a powerful computation substrate. We hope future works build upon our comprehensive study to better characterize{,} understand{, and enhance} the computational capability of DRAM chips.

\setcounter{version}{1}
\section*{\ieycr{0}{Acknowledgements}}
\ieycr{0}{We thank the anonymous reviewers of DSN 2024 for their encouraging feedback. 
We thank the SAFARI Research Group members for providing a stimulating intellectual environment. We acknowledge the generous gifts from our industrial partners, including Google, Huawei, Intel, and Microsoft. This work is supported in part by the Semiconductor Research Corporation (SRC), the ETH Future Computing Laboratory (EFCL), and the AI Chip Center for Emerging Smart Systems (ACCESS).}

\balance 
{
  \bstctlcite{IEEEexample:BSTcontrol}
  \let\OLDthebibliography\thebibliography
  \renewcommand\thebibliography[1]{
    \OLDthebibliography{#1}
    \setlength{\parskip}{0pt}
    \setlength{\itemsep}{0pt}
  }
  \bibliographystyle{IEEEtran}
  \bibliography{refs}
}

\onecolumn
\appendix
\section{Tested DRAM Modules}
\label{sec:appendix_tested_dram_modules}
\newcommand*{\myalign}[2]{\multicolumn{1}{#1}{#2}}
\newcommand{\dimmid}[2]{\begin{tabular}[l]{@{}l@{}}#1~\cite{datasheet#1}\\#2~\cite{datasheet#2}\end{tabular}}

Table~\ref{tab:detailed_info} shows the characteristics of the DDR4 DRAM modules we test and analyze. We provide {the} module and chip identifiers, access frequency (Freq.), manufacturing date (Mfr. Date), {chip} density (Chip Den.), die revision {(Die Rev.)}, chip organization {(Chip Org.), and the subarray size} of tested DRAM modules. We report the manufacturing date of these modules in the form of $week-year$. 
\renewcommand{\arraystretch}{1.2}
\begin{table*}[ht]
\footnotesize
\centering
\caption{Characteristics of the tested DDR4 DRAM modules.}
\label{tab:detailed_info}
\begin{tabular}{|l|l|l|c||c|cccc|c|}
\hline
\textbf{Module} & \textbf{Chip} & \textbf{Module Identifier} & \textbf{\#Modules} & \textbf{Freq} & \textbf{Mfr. Date} & \textbf{Chip} & \textbf{Die} & \textbf{Chip} & \textbf{Subarray} \\ 
\textbf{Vendor}&\textbf{Vendor}&\textbf{Chip Identifier}& \textbf{(\#Chips)} & \textbf{(MT/s)}&\textbf{ww-yy}&\textbf{Den.}&\textbf{Rev.}&\textbf{Org.}&\textbf{Size}\\
\hline
\hline
TimeTec & SK Hynix & \dimmid{TLRD44G2666HC18F-SBK}{H5AN4G8NMFR-TFC} & 7 (56) & 2666 & Unknown & 4Gb & M & $\times8$ & 512 or 640 \\
\cline{1-10}
TeamGroup & SK Hynix & {\dimmid{76TT21NUS1R8-4G}{H5AN4G8NAFR-TFC}} & 5 (40) & 2133 & Unknown & 4Gb & M & $\times8$ & 512 \\
\cline{1-10}
\hline
\hline
Micron & Micron & \dimmid{MTA4ATF1G64HZ-3G2E1}{MT40A1G16KD-062E:E} & 4 (16) & 3200 & 46-20 & 16Gb & E & $\times16$ & 1024 \\
\cline{1-10}
Micron & Micron & {\dimmid{MTA4ATF1G64HZ-3G2B2}{MT40A1G16RC-062E:B}} & 2 (8) & 2666 & 26-21 & 16Gb & B & $\times16$ & 1024 \\
\cline{1-10}
\hline 
\end{tabular}
\end{table*}

\end{document}